\documentclass[twocolumn,twocolappendix]{aastex63}
 
\newcommand{\msun}{$M_{\sun}$}
\newcommand{\rsun}{$R_{\sun}$}
\newcommand{\rstar}{$R_{\star}$}
\newcommand{\lsun}{$L_{\sun}$}
\newcommand{\msunyr}{\msun\,yr$^{-1}$}
\newcommand{\halpha}{H$\alpha$}

\newcommand{\mdot}{$\dot{M}$}
\newcommand{\ri}{R$_{\rm i}$}
\newcommand{\rw}{W$_{\rm r}$}
\newcommand{\tmax}{T$_{\rm max}$}

\newcommand{\chisq}{$\chi^2$}

\usepackage{graphicx}
\usepackage[caption=false]{subfig}
\usepackage{url}
\usepackage{hyperref}
\usepackage{comment}
\usepackage{amsmath}
\usepackage{booktabs}
\usepackage[version=4]{mhchem}

\usepackage{colortbl}
\usepackage{color}

\newcounter{column_number}
\shortauthors{Volz et al.}
\shorttitle{Carbon-rich Chemistry in a Transitional Disk}

\begin{document}

\title{JWST Reveals Carbon-rich Chemistry in a Transitional Disk}



\correspondingauthor{M. Volz}
\email{mvolz@bu.edu}

\author[0009-0005-4517-4463]{M. Volz}
\affil{Department of Astronomy, Boston University, 725 Commonwealth Avenue, Boston, MA 02215, USA}
\affil{Institute for Astrophysical Research, Boston University, 725 Commonwealth Avenue, Boston, MA 02215, USA}

\author[0000-0001-9227-5949]{C. C. Espaillat}
\affil{Department of Astronomy, Boston University, 725 Commonwealth Avenue, Boston, MA 02215, USA}
\affil{Institute for Astrophysical Research, Boston University, 725 Commonwealth Avenue, Boston, MA 02215, USA}

\author[0000-0001-9301-6252]{C. V. Pittman}
\affil{Department of Astronomy, Boston University, 725 Commonwealth Avenue, Boston, MA 02215, USA}
\affil{Institute for Astrophysical Research, Boston University, 725 Commonwealth Avenue, Boston, MA 02215, USA}

\author[0000-0002-4022-4899]{S. L. Grant}
\affil{Earth and Planets Laboratory, Carnegie Institution for Science, 5241 Broad Branch Road NW, 20015, Washington, DC, USA}

\author[0000-0003-4507-1710]{T. Thanathibodee}
\affil{Department of Physics, Faculty of Science, Chulalongkorn University, 254 Phayathai Road, Pathumwan, Bangkok 10330, Thailand}

\author[0000-0003-1878-327X]{M. McClure} 
\affil{Leiden Observatory, Leiden University, PO Box 9513, NL–2300 RA Leiden, The Netherlands}

\author[0000-0002-1103-3225]{B. Tabone}
\affil{Université Paris-Saclay, CNRS, Institut d'Astrophysique Spatiale, 91405, Orsay, France}
\affil{Leiden Observatory, Leiden University, PO Box 9513, NL–2300 RA Leiden, The Netherlands}

\author[0000-0002-3950-5386]{N. Calvet}
\affil{Department of Astronomy, University of Michigan, 1085 South University Avenue, Ann Arbor, MI 48109, USA} 

\author[0000-0001-7796-1756]{F. M. Walter}
\affiliation{Department of Physics and Astronomy, Stony Brook University, Stony Brook NY 11794-3800, USA}


\begin{abstract} 

We present JWST-MIRI Medium Resolution Spectrometer (MRS) observations of the Classical T Tauri stars GM~Aur and RX J1615.3-3255 (J1615), both hosting transitional disks. Despite their similar stellar and disk properties, the two systems differ strikingly in their carbon-bearing molecular emission. Using local thermodynamic equilibrium (LTE) slab models to analyze spectral lines within the $13.6$–$17.7$$\mu$m wavelength range, we find that J1615 exhibits strong emission from H$_2$O, HCN, C$_2$H$_2$, $^{12}$CO$_2$, $^{13}$CO$_2$, OH, and $^{13}$C$^{12}$CH$_2$, whereas GM~Aur shows only H$_2$O and OH. We measure the accretion rates of both objects using contemporaneous optical spectra and find that J1615's accretion rate is lower than that of GM~Aur. We constrain the properties of the dust in both disks using SED modeling and find elevated amounts of crystalline silicates and larger dust grains in the disk of J1615. The enhanced carbon emission in J1615 may result from a combination of lower accretion rate and larger and more processed dust grains in the inner disk, conditions that together may allow carbon-rich gas to persist and be detected. These results expand the sample of protoplanetary disks around solar-mass stars with strong CO$_2$ and C$_2$H$_2$ emission and identify J1615 as a carbon-rich transitional disk, providing new insights into the chemical diversity of planet-forming environments.

\end{abstract}




\section{Introduction} \label{intro}

The inner region ($<$10 au) of protoplanetary disks is a primary site of planet formation, containing the warm molecular gas and dust from which planets emerge \citep{Oberg-Bergin-2021, Dawson-Johnson-2018}. The molecular composition, temperature, and density in this chemically active environment determine the types of species available for forming planets \citep{Pontoppidan-etal-2014}, influencing their bulk composition, atmospheric properties, and potential habitability. Understanding these parameters is therefore fundamental to tracing the origin and diversity of planetary systems.

The \textit{Spitzer} Space Telescope provided the most detailed mid-infrared (MIR) view of protoplanetary disks before the launch of JWST. It revealed diverse molecular emission—H$_2$O, CO$_2$, HCN, and C$_2$H$_2$—but did not detect fainter species like $^{13}$CO$_2$ \citep{Pontoppidan-etal-2010, Salyk-etal-2011}. JWST’s higher spectral resolution now enables the detection of these weaker features \citep[e.g.,][]{Pontoppidan-etal-2024, Henning-etal-2024}. 

With \textit{Spitzer}, \cite{Pontoppidan-etal-2010} identified a class of ``CO$_2$-only" disks, exhibiting strong 14.98-$\mu$m CO$_2$ \textit{Q}-branch emission. JWST’s improved sensitivity has expanded the category to include disks with strong CO$_2$ and detectable, though weak, H$_2$O emission, as well as features from HCN, C$_2$H$_2$, $^{12}$CO$_2$, $^{13}$CO$_2$, OH, and $^{13}$C$^{12}$CH$_2$. Examples of ``CO$_2$-rich" disks with prominent CO$_2$ \textit{Q}-branch emission lines recently observed by JWST include GW Lup \citep[][]{Grant-etal-2023}, 
CX Tau \citep[][]{Vlasblom-2024-CXTau}, MY Lup \citep[][]{Salyk-etal-2025}, and XUE 10 \citep[][]{Frediani-etal-2025}.  

Beyond ``CO$_2$-rich" sources, \textit{Spitzer} data revealed a chemical dichotomy between cool stars/brown dwarfs (M5-M9) and young solar-mass (K1-M5) stars, with the former exhibiting more molecular carbon emission as evidenced by higher C$_2$H$_2$/HCN and HCN/H$_2$O flux ratios \citep{Pascucci-etal-2009, Pascucci-etal-2013}. Some JWST observations reinforce this trend. For instance, the low-mass stars ISO-ChaI 147 \citep[0.11 \msun,][]{Arabhavi-etal-2024} and 2MASS J16053215-1933159 \citep[0.14 \msun,][]{Tabone-etal-2023} feature emission from C$_2$H$_2$, $^{13}$C$^{12}$CH$_2$, and C$_4$H$_2$. Survey results from \cite{Arabhavi-etal-2025} and \cite{Grant-etal-2025} on the JWST-MIRI spectra of disks around very low mass stars (VLMS) confirm their hydrocarbon-rich nature. However, recent JWST observations have shown that carbon-rich disks can exist around solar-mass T Tauri stars \citep[e.g., DoAr 33, $1.1$ \msun,][]{Colmenares-etal-2024}.

The carbon-rich disks noted above are ``full" protoplanetary disks without any known large ($>10$ au) gaps. In this work, we present JWST-MIRI MRS spectra of the transitional disks of GM Aur and RX J1615.3-3255 (hereafter J1615).  Transitional disk SEDs are identified by deficits in disk emission in the NIR (1-5 $\mu$m) and MIR (5-20 $\mu$m), and excesses similar to that seen in full disks beyond $\sim 20 \mu$m which has been attributed to a cavity in the dust grain distribution. This cavity can extend from the central star to $\sim 10$s of au \citep{Espaillat-etal-2014}. 

The MIR molecular chemistry of transitional disks is not yet well explored, with most large-scale observational surveys including a handful of transitional disks among larger populations of full disks. The transitional disks around T Tauri stars with published JWST-MIRI spectra to date include 
SZ Cha \citep{Espaillat-etal-2023}, 
T Cha \citep{Xie-etal-2025}, 
UX Tau A \citep{Espaillat-etal-2024}, 
TW Hya \citep{Henning-etal-2024}, 
SY Cha \citep{Schwarz-etal-2024}, 
PDS 70 \citep{Perotti-etal-2023}, 
MY Lup \citep{Salyk-etal-2025}, 
HP Tau \citep{Romero-Mirza-etal-2024b},
RY Lup, 
AS 205 S,
DoAr 25,
IRAS 04385+2550,
and SR 4 \citep{JDISC-2025}.

\begin{table*}[]
\hspace{-5em}
\begin{tabular}{lllllllll}
\hline
Target & d [pc] & Spectral type & $R_*$ [\rsun] & $T_{\rm eff}$ [K] & $L_*$ [\lsun] & $A_{V}$ & M$_{*}$ [\msun] & $\log$($\dot{M}$ [\msunyr]) \\
\hline
\hline

GM Aur & $155^{+1.4}_{-2.0}$$^{(a)}$ & K5$^{(b)}$ & $1.75 \pm 0.51^{(b)}$ & $4350^{(b)}$ & $1.29^{(c)}$ & $0.6^{(b)}$ & $1.36 \pm 0.36^{(b)}$ & $-7.7^{(d)}$ \\

RXJ1615.3-3255 & $156^{+0.57}_{-0.57}$$^{(a)}$ & K7$^{(b)}$ & $1.90 \pm 0.55^{(b)}$ & $4060^{(b)}$ & $0.63^{(c)}$ & $0.0^{(b)}$ & $1.16 \pm 0.16^{(b)}$ & $-8.6^{(d)}$ \\

\hline
\end{tabular}
\caption{A summary of the stellar parameters for the two objects discussed in this paper. (a) \cite{GaiaCollaboration-etal-2023}; (b) \cite{Manara-etal-2014}; (c) These values, originally from \cite{Manara-etal-2014}, were scaled in accordance with the new Gaia distances; (d) This work (see Section \ref{acc}).}
\label{table:stellar-parameters}
\end{table*}

The observations of GM Aur and J1615 presented in this work offer a unique opportunity to compare and contrast the molecular chemistry of two otherwise similar transitional disk systems. GM Aur is a K5 CTTS star with an accretion rate of $\log(\dot{M}) \approx -7.7$ \msunyr  and $M_* = 1.36 \pm 0.36$ \msun (\citealt{Manara-etal-2014}; see Table \ref{table:stellar-parameters}) with a transitional disk located 155 pc away in Taurus-Auriga \citep{GaiaCollaboration-etal-2023}. ALMA and VLA millimeter imaging reveal a $\sim$35 au inner cavity with a small ($<3.2$ au) inner disk \citep{Macias-etal-2018, Francis-vanderMarel-2020}. Located 156 pc away in the Lupus star-forming region, J1615 hosts a transitional disk \citep{Merin-etal-2010} around a K7 CTTS with an accretion rate of $\log(\dot{M}) \approx -8.6$ \msunyr and $M_* = 1.16 \pm 0.16$ \msun (\citealt{Manara-etal-2014}). ALMA millimeter dust imaging shows a $\sim$30 au cavity with an inner disk that may extend up to $\sim 15$ au \citep{Sierra-etal-2024}. 

In Section \ref{redux}, we present the observations of GM Aur and J1615. In Section \ref{analysis}, we use slab modeling to analyze the molecular emission lines of H$_2$O, HCN, C$_2$H$_2$, $^{12}$CO$_2$, $^{13}$CO$_2$, OH, and $^{13}$C$^{12}$CH$_2$. We also explore the disk properties of both targets by modeling their spectral energy distributions (SEDs). We measure MIR atomic emission line fluxes for [\ion{Ne}{2}], [\ion{Ne}{3}], and [\ion{Ar}{2}] and calculate accretion rates from contemporaneous optical spectra. We also compare the new JWST spectra to archival \textit{Spitzer} spectra to search for MIR continuum variability. In Section \ref{disc} we discuss the possible mechanisms behind the differences in the carbon chemistry of J1615 and GM Aur. We end with a summary in Section \ref{conclusions}.


\section{Observations \& Data Reduction} \label{redux}

Here we present new JWST Mid-InfraRed Instrument (MIRI) spectra of GM~Aur and J1615 along with new high-resolution Chiron optical spectra that were taken contemporaneously (within $\sim12$ hrs) of the JWST observations.


\subsection{MIR spectra} \label{jwst} \label{MIRdata}

GM Aur and J1615 were observed with JWST MIRI \citep{Rieke-etal-2015, Wells-etal-2015, Wright-etal-2015, Wright-etal-2023, Argyriou-etal-2023} in its Medium Resolution Spectroscopy (MRS) mode on September 27 and August 25, 2023, respectively, as part of JWST General Observers Program 1676 (PI: Espaillat). Both targets had their own separate background observations. Observation details are summarized in Table \ref{table:jwstobs}.

\begin{figure*}  
\epsscale{1.2}
\includegraphics[width = 1.0\textwidth]{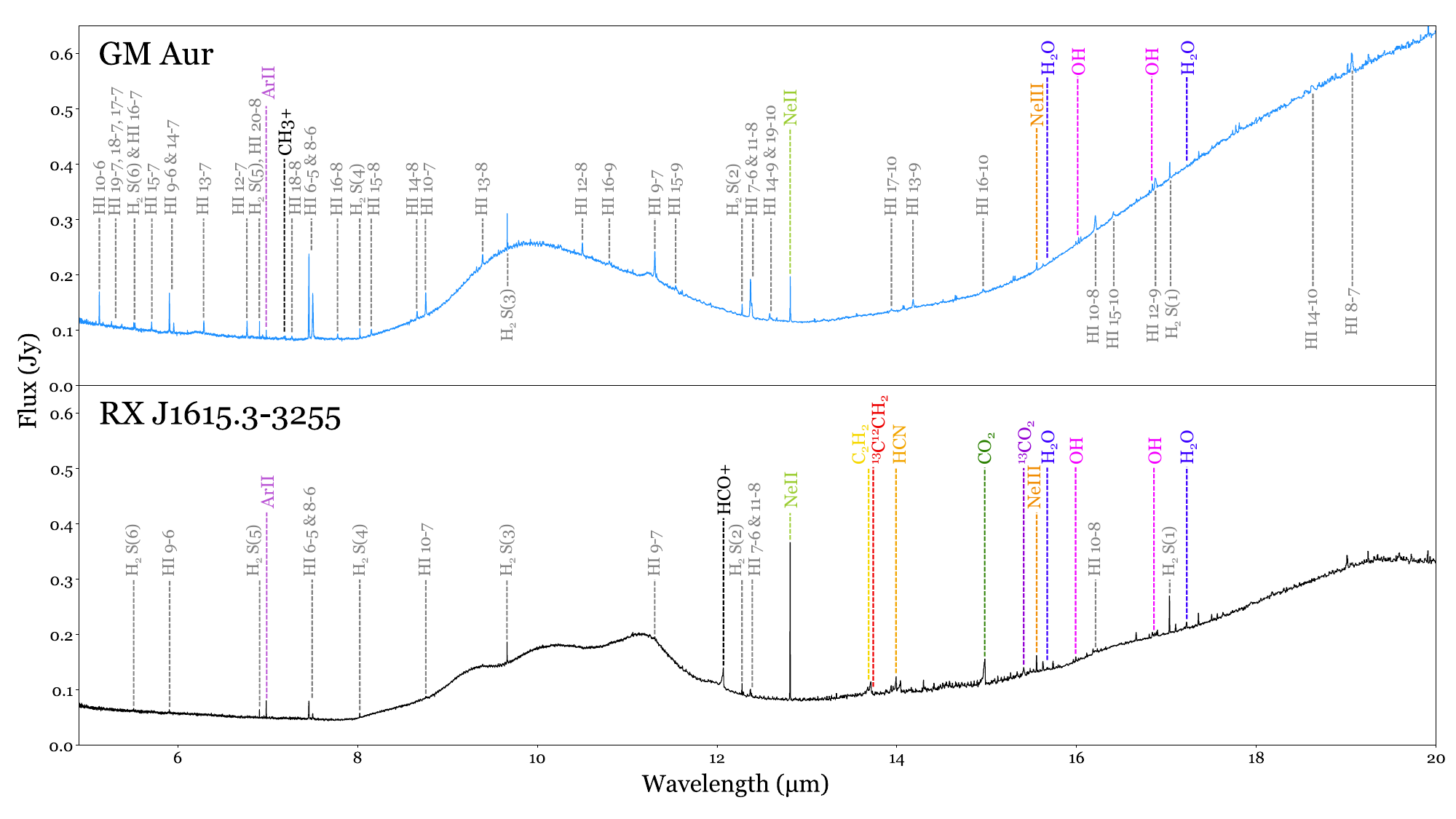}
\caption{
The JWST-MIRI spectra for transitional disks GM Aur (\textit{blue}) and J1615 (\textit{black}). Bright atomic and molecular hydrogen lines are labeled with grey dashed lines. Molecular ions are labeled in black. Atomic and molecular features  analyzed in this work are denoted with colored text. Wavelengths beyond 20 $\mu$m are omitted, but are shown in Figure \ref{fig:variability}.
}
\label{fig:fullspec}
\end{figure*} 

\begin{table*}[]
\begin{tabular}{lllll}
\hline
Target          & Start time (UT)          & End time (UT)          & Exposure time [s] & Observing time [hr] \\
\hline
\hline
GM Aur          & 2023-09-27 21:16:54 & 2023-09-27 22:24:00 & 27.750              & 1.46           \\
RX J1615.3-3255 & 2023-08-25 04:07:14 & 2023-08-25 02:55:36 & 27.750              & 1.46 \\
\hline
\end{tabular}
\caption{Descriptions of JWST MIRI-MRS observations. The exposure times are per sub-band exposure, and the observing time refers to the total amount of time spent on target.}
\label{table:jwstobs}
\end{table*}

The MIRI observations are processed through all three stages of the JWST Reduction Pipeline version 1.19.2 \citep{Bushouse-etal-2024}. We used reference file \texttt{jwst\_1413.pmap} and followed the same procedure as in \cite{Espaillat-etal-2023}. The reduced MIRI-MRS spectra of GM Aur and J1615 are presented in Figure \ref{fig:fullspec}. Both targets have detections of [\ion{Ar}{2}] at 6.98 $\mu$m, [\ion{Ne}{2}] at 12.81 $\mu$m, and [\ion{Ne}{3}] at 15.5 $\mu$m. In GM Aur, numerous \ion{H}{1} and H$_2$ emission lines are present, along with OH and H$_2$O while J1615 shows several molecular features in the $\sim13.6 – 17.7$ $\mu$m range including H$_2$O, HCN, C$_2$H$_2$, $^{12}$CO$_2$, $^{13}$CO$_2$, OH, and $^{13}$C$^{12}$CH$_2$ (see Figure \ref{fig:fullspec}). 


\subsection{Optical Spectra} \label{optical}

GM~Aur and J1615 were observed with the Chiron echelle spectrograph on the SMARTS/CTIO 1.5~m telescope \citep{Tokovinin2013} contemporaneously with their JWST observations. GM~Aur was observed three times during 28--30 September 2023, with the closest observation being 11.6 hours after JWST. J1615 was observed five times during 23--27 August 2023, with the closest observation being 2.4 hours before JWST (see Tables \ref{table:jwstobs} and \ref{table:accretion-parameters}). These observations provided the \halpha\ profiles needed to measure their accretion rates. The spectra were obtained in fiber mode with $4\times4$ on-chip binning, which produced a spectral resolution of $R\sim27,800$. The data are reduced using a custom IDL pipeline.\footnote{\url{http://www.astro.sunysb.edu/fwalter/SMARTS/CHIRON/ch_reduce.pdf}} They are then corrected for radial velocity and photospheric absorption using a PHOENIX model spectrum \citep{husser13} with solar metallicity and stellar $T_{\rm eff}$ and $\log g$ (assuming the parameters in Table~\ref{table:stellar-parameters}).  The data are presented and modeled in Section~\ref{acc}.


\section{Analysis \& Results} \label{analysis}

Here we analyze the gas and dust properties of GM~Aur and J1615.  We perform slab modeling of the molecular gas emission from H$_2$O, HCN, C$_2$H$_2$, $^{12}$CO$_2$, $^{13}$CO$_2$, OH, and $^{13}$C$^{12}$CH$_2$.  We also measure atomic line fluxes. Then we model the {\halpha} profiles obtained contemporaneously with Chiron to measure accretion rates and properties of the accretion flow. We also use SED models to constrain the the dust properties of both targets' inner and outer disks. Lastly, we compare the JWST spectra with archival \textit{Spitzer} spectra to search for variability in the dust continuum.


\subsection{Molecular Gas Emission Lines} \label{slabmodels}

We fit the $13.6 – 17.7$ $\mu$m range because it contains bright emission from multiple species and it is where the stark difference in molecular brightness between J1615 and GM Aur is most striking. In J1615, features from CO$_2$'s \textit{Q}-branch, hot bands (at $13.85$ and $16.2$ $\mu$m), and its isotopologue $^{13}$CO$_{2}$ are visible alongside bright C$_2$H$_2$ and HCN features at shorter wavelengths, and some H$_2$O lines at longer wavelengths. In GM Aur, bright OH features and some water is visible among bright \ion{H}{1} and H$_2$ lines.

We first continuum subtract each spectrum. We use the package \texttt{ctool} \footnote{\url{https://github.com/pontoppi/ctool}}, which determines the continuum by first masking any data in the original spectrum above a given threshold. Next, a median smoothing filter (SciPy’s \texttt{medfilt}; \citealt{Virtanen-etal-2020}) is applied to the desired sections with boxsize defined by the user (in our case, 95), creating a smoothed spectrum. The continuum level is set at points where the smoothed spectrum is above 99.8\% of the original spectrum. These points are removed, leaving a masked spectrum. The original spectrum is then interpolated with the masked spectrum five times. Then, a second-order Savitzky-Golay filter with window length equal to three times the boxsize is applied, generating a filtered spectrum. Finally, this continuum level is subtracted from the filtered spectrum. Thus, we produce the continuum-subtracted data.

We fit the data with a slab model presented in \cite{Tabone-etal-2023} and used in other works \citep[e.g.,][]{Perotti-etal-2023,Temmink-etal-2024,Ramirez-Tannus-etal-2023} and follow the fitting procedure as outlined in \cite{Grant-etal-2023}.  In short, the slab modeling routine compares the continuum-subtracted spectrum in a defined fitting range to sample emission spectra from various molecular species generated from HITRAN database line transitions. The parameters of the model (temperature, column density, and emitting area) are varied until the $\chi^2$ between the model and the observed data is minimized, thus providing best-fit values for each chemical species in the disk (See Appendix \ref{chi2s} for more details).

The fit begins with a Gaussian line profile of FWHM $\Delta V$ = 4.7 km s$^{-1}$, representative of H$_2$ at 700 K. The line model accounts for mutual shielding of lines for the same species, and has three free parameters: temperature ($T$), line-of-sight column density ($N$), and emitting area ($\pi R^2$). While the emitting area is parametrized by a full disk with no inner hole, it should be noted that the molecular emission could also be described by a ring with the same area. 

We fit emission from the molecules H$_2$O, HCN, C$_2$H$_2$, $^{12}$CO$_2$, $^{13}$CO$_2$, OH, and $^{13}$C$^{12}$CH$_2$. For each molecule, the code iterates through a grid of $N$ and $T$ values. For $N$, this grid is between $10^{14}$ and $10^{22}$ cm$^{-2}$ with steps of 0.166 in log$_{10}$-space. For $T$, the grid is defined between 100\,K to 1500\,K with steps of 25 K. The emitting radius $R$ is varied in steps of 0.03 in log$_{10}$-space between 10$^{-2}$ au and 10 au. A model spectrum is generated with each parameter value, convolved to MIRI’s spectral resolution \citep[$R = 2500$,][]{Labiano-etal-2021}, and compared to the observed spectrum. For each $N$ and $T$, the emitting area $\pi R^2$ is varied until the reduced $\chi^{2}$ is minimized. We use the same $\chi^2$ calculation method as \cite{Grant-etal-2023}, presenting the $\chi^2$ maps for each target in Figure~\ref{fig:chi2s} in Appendix \ref{chi2s}.

\begin{figure*}
\centering
\includegraphics[width = 1.0\textwidth]{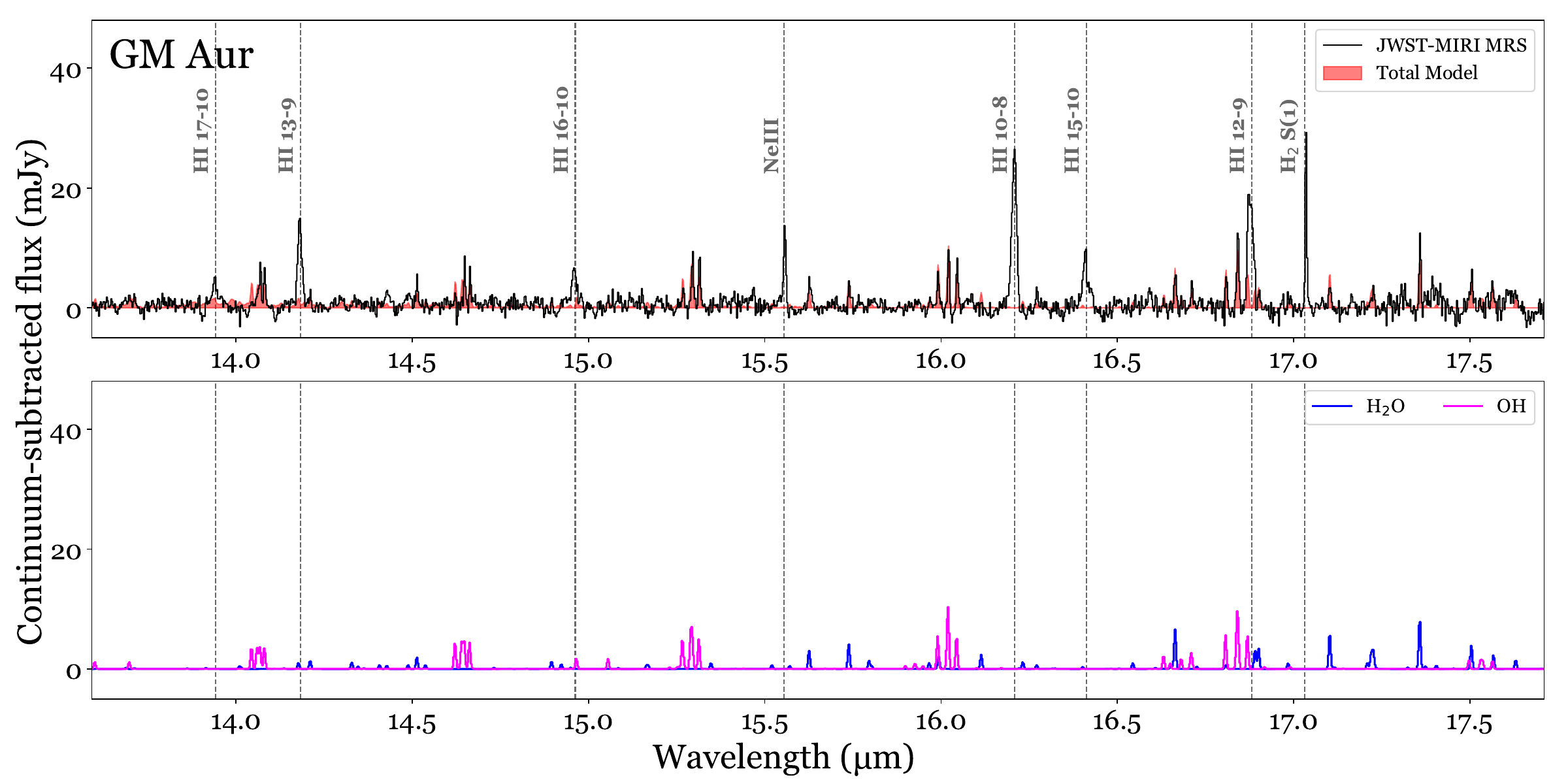}
\caption{The $13.6-17.7$ $\mu$m range of the GM Aur spectrum. The continuum-subtracted JWST-MIRI data (\textit{black}, top panel) is overlaid with the total modeled emission (\textit{red}, top panel) from molecules (bottom panel) H$_2$O (\textit{blue}) and OH (\textit{pink}). Slab model fits were also performed for HCN, C$_2$H$_2$, $^{12}$CO$_2$, $^{13}$CO$_2$, and $^{13}$C$^{12}$CH$_{2}$, however, H$_2$O and OH were the only positive detections, while the remaining species did not match our detection criteria (see Section \ref{slabmodels}). The detected bright \ion{H}{1}, H$_2$, and atomic emission lines are labeled with dashed lines.}
\label{fig:gmaur-fit}
\includegraphics[width = 1.0\textwidth]{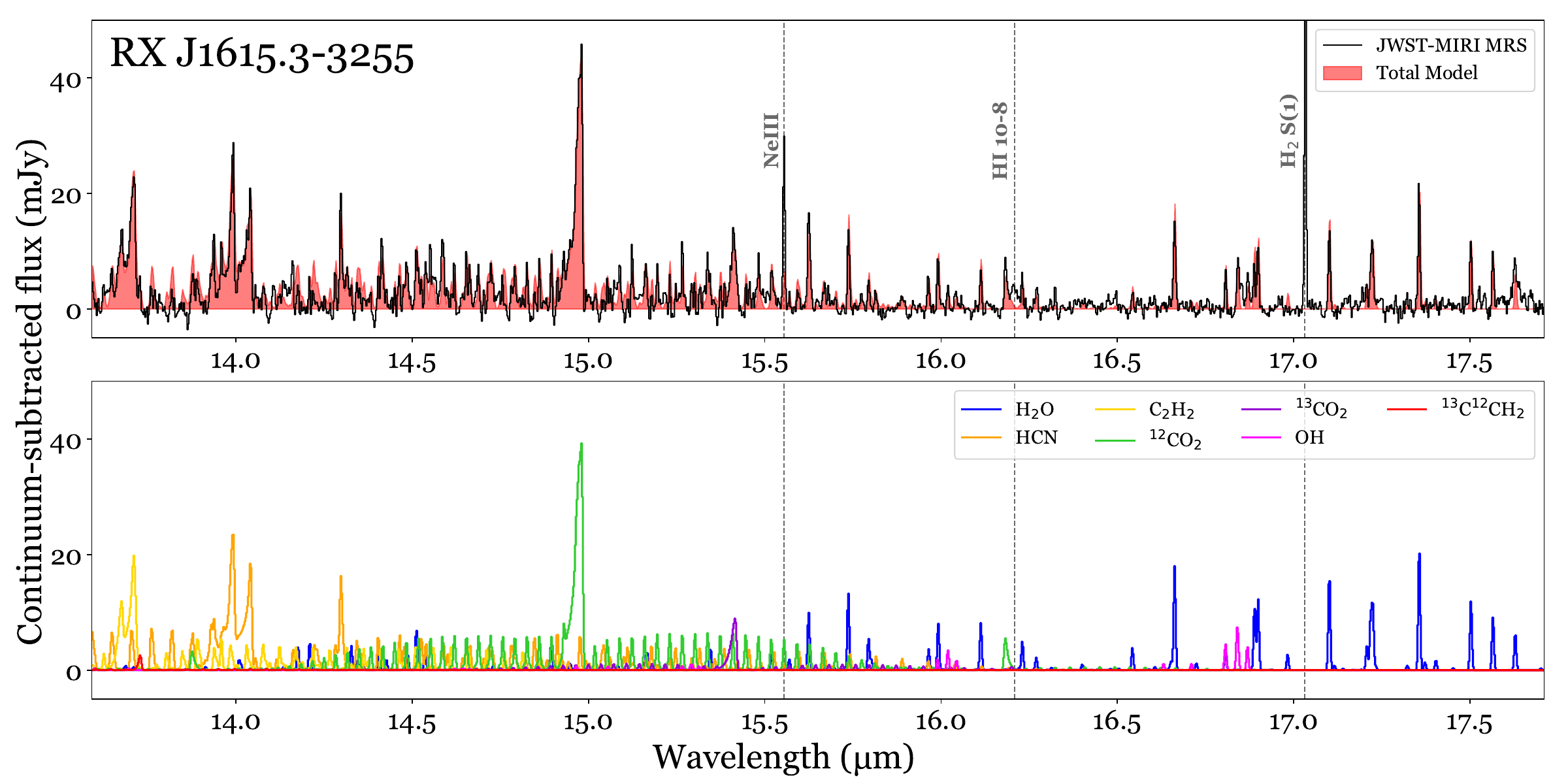}
\caption{
The $13.6 – 17.7$ $\mu$m wavelength range of the J1615 spectrum. The continuum-subtracted JWST-MIRI data (\textit{black}, top panel) is overlaid with the total modeled emission (\textit{red}, top panel) from the component molecules (bottom panel): H$_2$O (\textit{blue}), HCN (\textit{orange}), C$_2$H$_2$ (\textit{yellow}), $^{12}$CO$_2$ (\textit{green}), $^{13}$CO$_2$ (\textit{purple}), OH (\textit{pink}), $^{13}$C$^{12}$CH$_{2}$ (\textit{red}). Based on our detection criteria (see Section \ref{slabmodels}), we consider all fitted species positive detections. The detected bright \ion{H}{1}, H$_2$, and atomic emission lines are labeled with dashed lines.
}
\label{fig:RXJslabs}
\end{figure*}

The slab modeling code returns best-fit parameters for molecules H$_2$O, C$_2$H$_2$, HCN, $^{12}$CO$_2$, $^{13}$CO$_2$, OH, and $^{13}$C$^{12}$CH$_2$, fit in that order in J1615. We fit H$_2$O first and OH second in GM Aur as they are the brightest species. The final best-fit molecular spectra are shown alongside the continuum-subtracted JWST-MIRI spectrum for GM Aur and J1615 in Figure~\ref{fig:gmaur-fit} and Figure~\ref{fig:RXJslabs}, respectively. 

The detection or non-detection of a molecular species is determined separately to the best-fit slab model parameters and their associated \chisq. In order to determine the presence of each molecule, we first subtract the best-fit slab models for all other species except for one molecule of interest. In the resultant spectrum, we determine the signal-to-noise ratio (SNR) of the peak of the brightest emission feature. If the SNR is above 10, we consider it a positive detection. If SNR $<10$, it is a non-detection.

The best-fit parameters for GM Aur and J1615 are summarized in Table \ref{table:bestfit} and \ref{table:bestfitGM}. In the subsections that follow, we report the best-fit slab model parameters and detection status for each molecule.

\begin{table}[]
\hspace{-1em}
\begin{tabular}{llll}
\hline
Species               & N (cm$^{-2}$)          & T (K) & R (AU) \\
\hline
\hline
H$_2$O                & 1.00 $\times$ 10$^{18}$  & 450   & 0.32 \\
HCN                   & 4.64 $\times$ 10$^{19}$  & 225   & 0.28 \\
C$_2$H$_2$            & 3.16 $\times$ 10$^{20}$  & 100   & 4.32 \\
$^{12}$CO$_2$         & 2.15 $\times$ 10$^{18}$  & 200   & 0.50 \\
$^{13}$CO$_2$         & 4.64 $\times$ 10$^{16}$  & 275   & 0.22 \\
OH                    & 3.16 $\times$ 10$^{16}$  & 850   & 0.40 \\
$^{13}$C$^{12}$CH$_2$ & 2.15 $\times$ 10$^{14}$  & 100   & 9.32 \\
\hline
\end{tabular}
\caption{Slab model best-fit parameters for molecular emission in J1615. Note that when a molecule is fitted by an optically thin ($\lesssim 1 \times 10^{18}$ cm$^{-2}$) slab, there is a degeneracy between small $N$ and large $R$ and large $N$ and small $R$. The $^{13}$C$^{12}$CH$_2$ line list is incomplete, and thus the best-fit parameters should be taken with caution.}
\label{table:bestfit}
\end{table}

\begin{table}[h!]
\hspace{-1em}
\begin{tabular}{llll}
\hline
Species               & N (cm$^{-2}$)          & T (K) & R (AU) \\
\hline
\hline
H$_2$O                & 2.15 $\times$ 10$^{14}$  & 475   & 9.32 \\
OH                    & 4.64 $\times$ 10$^{18}$  & 1200   & 0.05 \\
\hline
\end{tabular}
\caption{Slab model best-fit parameters for molecular emission in GM Aur. We only list the parameters for the species with positive detections. Note that when a molecule is fitted by an optically thin ($\lesssim 1 \times 10^{18}$ cm$^{-2}$) slab, there is a degeneracy between small $N$ and large $R$ and large $N$ and small $R$. The $^{13}$C$^{12}$CH$_2$ line list is incomplete, and thus the best-fit parameters should be taken with caution.}
\label{table:bestfitGM}
\end{table}


\subsubsection{H$_2$O} \label{h2o}

Generally speaking, the spectra of GM Aur and J1615 appear more water-poor than the disks around many other T Tauri stars \citep{JDISC-2025, Temmink-etal-2025, Banzatti-etal-2023}. We perform a slab model fit of the water in J1615 and GM Aur to better understand its parameters. In J1615 (Figure \ref{fig:RXJslabs}), the brightest water lines are found past 15.5 $\mu$m and are best fit by a slab model of column density $1.00 \times 10^{18}$ cm$^{-2}$, temperature 450 K, and emitting radius 0.32 au. 
As for GM Aur, several water lines are visible in its spectrum between $15.5 - 16.0$ $\mu$m and $17.0 - 17.5$ $\mu$m. Our slab models best fit H$_2$O in GM Aur with a column density of 2.15 $\times$ 10$^{14}$ cm$^{-2}$, temperature 475 K, and emitting radius 9.32 au. We use the peak at $\sim17.358$ $\mu$m to determine an SNR of 10.3 in GM Aur and 23.2 in J1615, both positive detections.

It is now possible to go beyond a single temperature component when fitting water emission in MIR spectra \citep{Temmink-etal-2025, Banzatti-etal-2025}, especially when one component is ineffective in replicating the observed data (for example, \citealt{Grant-etal-2024}). We find that a one-component temperature slab model is an effective recreation of the water lines between $13.6 - 17.7$ $\mu$m in both GM Aur and J1615.

\begin{figure*}
    \includegraphics[width = 1.0\linewidth]{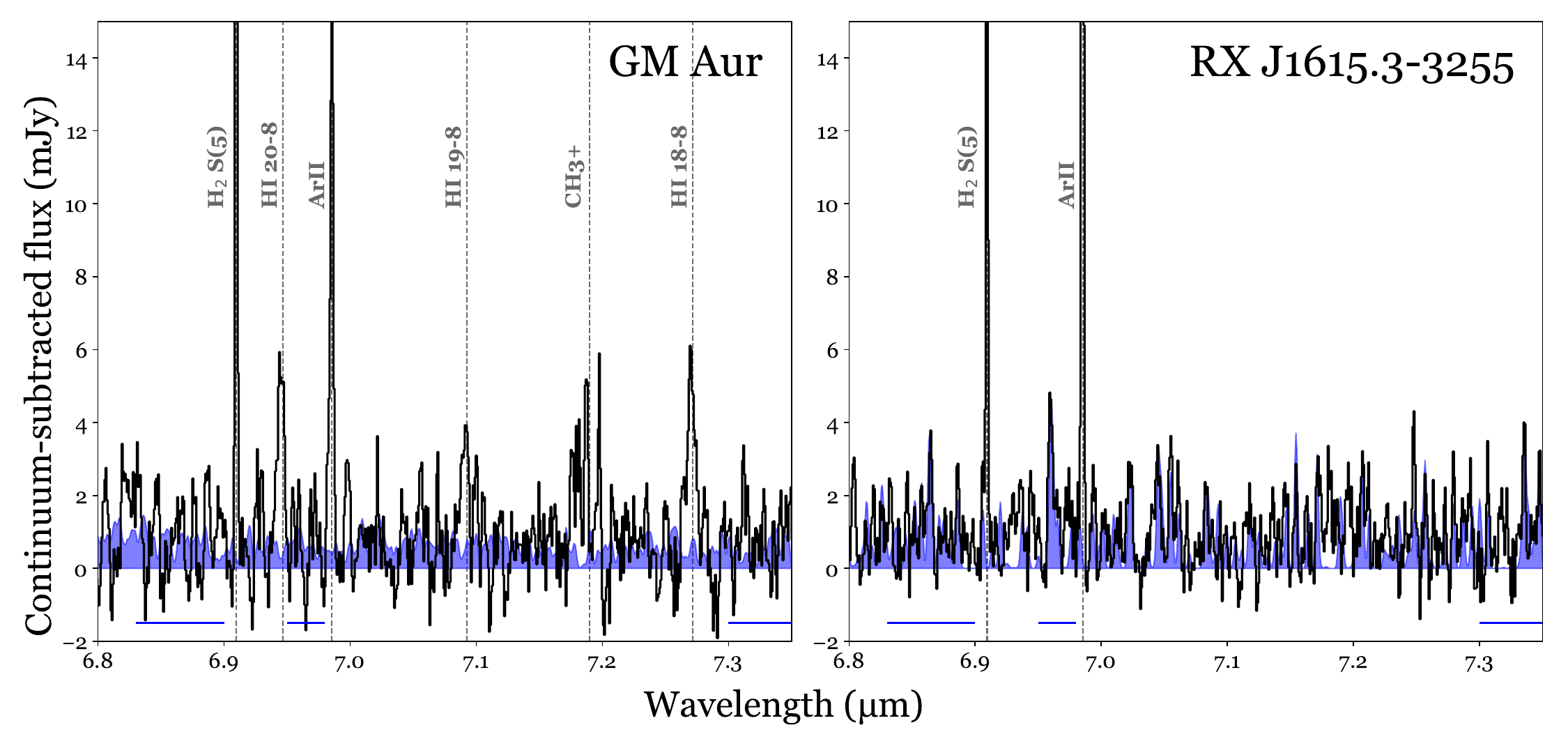}
    \caption{LTE slab model fits (\textit{blue}) of the $\sim6.8-7.5$ $\mu$m bending mode of water in the disks of GM Aur and J1615 (JWST-MIRI spectrum in \textit{black}). Bright atomic and molecular hydrogen and other atomic and ionic emission lines are labeled with grey dashed lines. The horizontal blue lines are the fitting windows used to calculate $\chi^2$ (see Appendix \ref{chi2s}). As stated in Section \ref{h2o}, the presence of water lines at these wavelengths in J1615 may indicate a water reservoir closer to the central star.}
    \label{fig:6-7um_slabs}
\end{figure*}

For transitional disks, the relative intensity between excited water lines at short wavelengths and less excited lines at longer wavelengths is particularly informative when we consider the impacts of inner gas and dust cavities. \cite{Banzatti-etal-2017} find that the relative line fluxes of water decrease from hot water at shorter wavelengths to cold water at longer wavelengths as the inner carbon monoxide radius ($R_{CO}$) increases. In other words, as the inner gas cavity of a disk grows, water is depleted from the inside out in the disk. We can demonstrate this effect in the JWST-MIRI spectra of GM Aur and J1615. In the $17.0 - 17.5$ $\mu$m wavelength range (visible in Figures \ref{fig:gmaur-fit} and \ref{fig:RXJslabs}), water lines are found in both disks, albeit brighter in J1615. In Figure \ref{fig:6-7um_slabs}, we present the $6.8-7.5$ $\mu$m region of the H$_2$O bending mode in both spectra. The observational data, in black, is overlaid with an LTE slab model fit to the water lines. At these wavelengths, water is detected in J1615 (see the bright line near $\sim6.96$ $\mu$m), while no water lines are seen in GM Aur. The non-detection of water in GM Aur is supported by the poor visual fit of its LTE slab model and a $\chi^2$ of 3.44 (vs. 0.96 for J1615, see Figure \ref{fig:chi2s_h2o} in the Appendix). 

In the disk of PDS 70, water lines between $6.8 - 7.5$ $\mu$m suggest that water vapor may persist in the disk’s terrestrial planet-forming regions \citep{Perotti-etal-2023}. The detection of water lines in this same wavelength range in J1615 and not in GM Aur may indicate that a water reservoir exists at smaller radii in J1615 and not in GM Aur; while both disks have water at larger radii, cooler temperatures, and longer wavelengths, only J1615 has maintained a reservoir closer to its central star. This is supported by the best-fit parameters for the $6.8 - 7.5$ $\mu$m slab model–$T = 775$ K, $N = 4.64 \times 10^{18}$ cm$^{-2}$, and $R = 0.028$ au–which are a hotter temperature and smaller emitting radius than the best-fit values for the longer wavelength H$_2$O lines.

\subsubsection{HCN} \label{hcn}

HCN appears prominently in J1615 as a triple-peaked feature at 14.0 $\mu$m. It is best reproduced with a slab of column density $4.64 \times 10^{19}$ cm$^{-2}$, temperature 225 K, and emitting radius 0.28 au. The SNR of its peak at $\sim13.99$ $\mu$m is 47, a very bright detection. We calculate the HCN/C$_2$H$_2$ flux ratio in J1615 by first subtracting the best-fit slab spectra for all modeled species except HCN and C$_2$H$_2$, then integrating along the wavelength region where HCN and C$_2$H$_2$ are expected to be ($13.62-13.73$ $\mu$m for C$_2$H$_2$ and $13.91-14.1$ $\mu$m for HCN). We find a flux ratio of $1.37 \pm 0.19$, in alignment with the young solar analogs presented in \cite{Pascucci-etal-2009}, which show higher HCN fluxes than low-mass stars. The HCN/C$_2$H$_2$ column density ratio of J1615 is 0.147. This is more aligned with brown dwarf disks as presented in \cite{Pascucci-etal-2013}.   

In GM Aur, the models for HCN are highly degenerate. The SNR at $\sim13.99$ $\mu$m is only 1.8, which qualifies as a non-detection.


\subsubsection{C$_2$H$_2$ \& $^{13}$C$^{12}$CH$_2$} \label{c2h2}

Both C$_2$H$_2$ and its isotopologue $^{13}$C$^{12}$CH$_2$ are observed in J1615. The former is best described with column density $3.16 \times 10^{20}$ cm$^{-2}$, temperature 100 K, and emitting radius 4.32 au. The latter is representative of a less dense ($N = 2.15 \times 10^{14}$ cm$^{-2}$), same temperature ($T = 100$ K), and more distant ($R = 9.32$ au) population. Thus, $^{13}$C$^{12}$CH$_2$ may originate from a deeper layer of the disk. We should note that the line list for $^{13}$C$^{12}$CH$_2$ is likely incomplete, and any best-fit values should be taken with caution. The strong C$_2$H$_2$ \textit{Q}-branch near 13.7 $\mu$m has an SNR of 34 and the nearby $^{13}$C$^{12}$CH$_2$ peak has an SNR of 12, both positive detections in J1615.

The same C$_2$H$_2$ and $^{13}$C$^{12}$CH$_2$ in GM Aur have SNRs of 3.2 and 0.5, respectively. Thus, neither C$_2$H$_2$ nor $^{13}$C$^{12}$CH$_2$ is detected in this wavelength range.


\subsubsection{$^{12}$CO$_2$ \& $^{13}$CO$_2$} \label{co2}

Perhaps the most recognizable feature in J1615's spectrum is the 14.98-$\mu$m \textit{Q}-branch of $^{12}$CO$_2$, observed with a peak SNR of 70. Its strong \textit{P}- and \textit{R}-branches frame the central feature, alongside ``hot bands" at $13.9$ and $16.2$ $\mu$m. The isotopologue $^{13}$CO$_2$ is also observed at $15.4$ $\mu$m. This is very similar to GW Lup \citep{Grant-etal-2023}, another ``CO$_2$-rich" source. As shown in their analysis, $^{12}$CO$_2$ must be optically thick to recreate the observed features.

$^{12}$CO$_2$ in J1615 is fit with a slab with column density $2.15 \times 10^{18}$ cm$^{-2}$, temperature 200 K, and emitting radius 0.50 au. $^{13}$CO$_2$'s slab model traces a column density of $4.64 \times 10^{16}$ cm$^{-2}$, temperature 275 K, and emitting radius 0.22 au. In other targets with positive $^{12}$CO$_2$ and $^{13}$CO$_2$ detections such as GW Lup \citep{Grant-etal-2023}, CX Tau \citep{Vlasblom-2024-CXTau}, and XUE 10 \citep{Frediani-etal-2025}, $^{13}$CO$_2$ also had a lower column density than $^{12}$CO$_2$, but unlike our results, was \textit{colder} than $^{12}$CO$_2$.  

In GM Aur, the slab models attempt a fit for $^{12}$CO$_2$, but the \textit{Q}-branch overlaps with \ion{H}{1} 16-10 and the $^{12}$CO$_2$ hot band at $16.2$ $\mu$m with \ion{H}{1} 10-8. We report a non-detection of $^{12}$CO$_2$ and $^{13}$CO$_2$ in GM Aur, both with peak SNRs below 4. 


\subsubsection{OH} \label{oh}

\begin{figure*}
    \centering
    \includegraphics[width=1.0\textwidth]{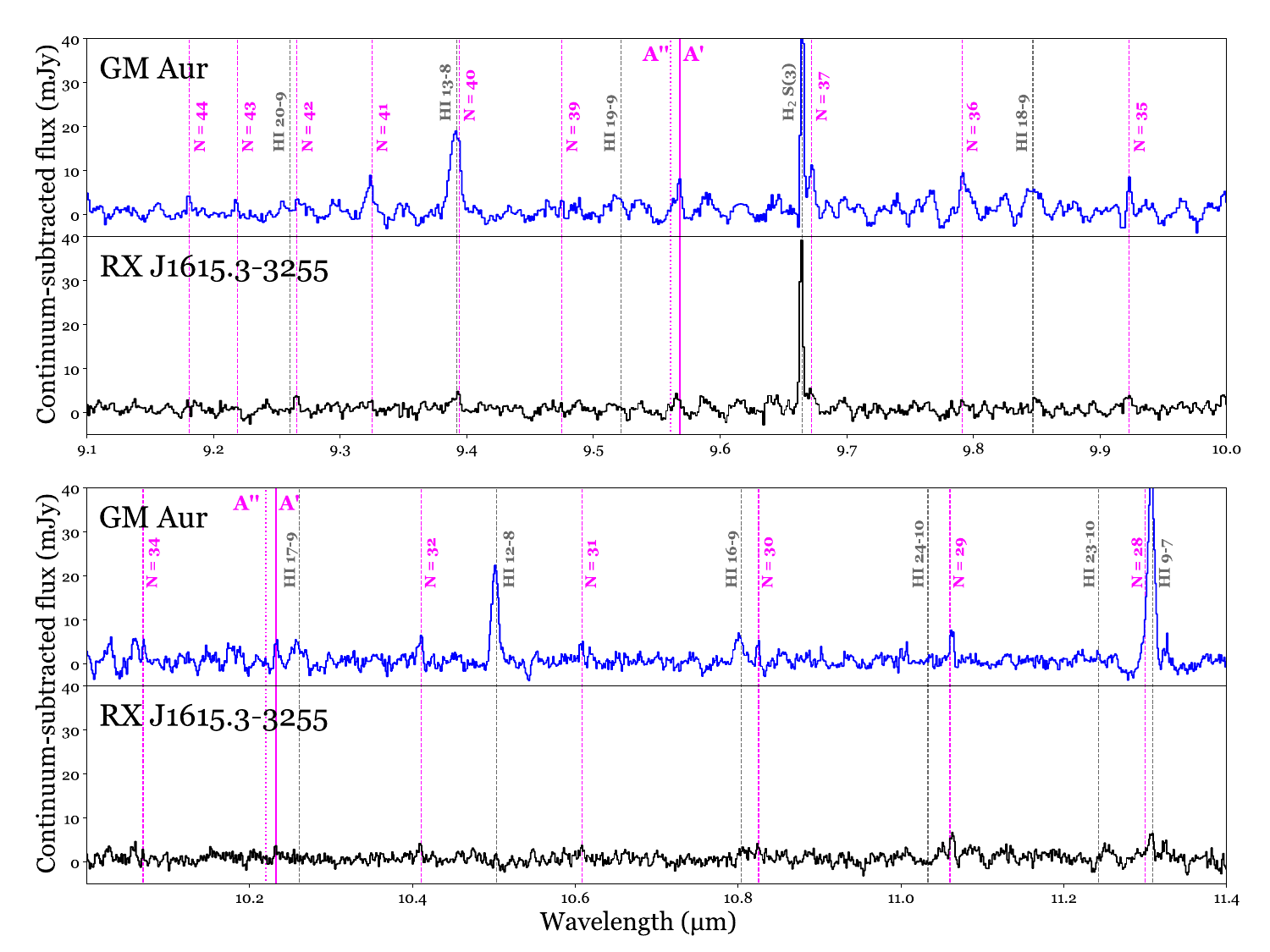}
    \caption{The $9 - 10$ $\mu$m and $10.1 - 11.5$ $\mu$m segments of the JWST-MIRI MRS spectra of GM Aur (\textit{blue}) and J1615 (\textit{black}). \ion{H}{1} and H$_2$ lines are labeled with vertical gray dashed lines. The location of OH emission lines are indicated with vertical pink lines alongside their associated quantum number. $A'$ and $A''$ symmetry lines are labeled, with the former being a strong indicator of prompt emission. The higher detection rate of OH lines in GM Aur over J1615 (see Section \ref{oh}) suggests that the photodissociation of H$_2$O into OH is more important in the disk of GM Aur than J1615.}
    \label{fig:OHlines}
\end{figure*}

We slab model OH emission within $13.6 - 17.7$ $\mu$m. In J1615, the OH is brightest near $\sim 16.8$ $\mu$m and has a peak SNR of 19. Its slab model fit, however, is highly degenerate. While we do report a positive detection of OH in J1615, we cannot make any strong constraints on its column density, temperature, or emitting radius.

GM Aur has multiple bright OH features between $13.6 - 17.7$ $\mu$m which peak with an SNR of 31. The best-fit parameters for our OH slab model are $N = 4.64 \times 10^{18}$ cm$^{-2}$, $T = 1200$ K, and $R = 0.05$ au. Similar to the findings of \cite{JDISC-2025}, we report a positive detection of OH in our GM Aur spectrum. 

In addition to OH features between $13.6 - 17.7$ $\mu$m, emission lines between $9 - 12$ $\mu$m are known to originate from highly excited OH in transition from rotational quantum numbers $N \rightarrow N-1$ beginning at $N = 44$. These states correlate with upper-level temperatures near $40000$ K. The photodissociation of H$_2$O by UV photons around Ly$\alpha$ is a driving mechanism that pushes OH into these high energy levels, and whose subsequent fall produces ``prompt emission" \citep{vanHarrevelt-vanHemert-2003}. We identify lines unique to H$_2$O dissociation by the fact that the resultant OH inhabits two of four hyperfine states, often labeled as the $A'$ symmetry and the other as $A''$ \citep{Zhou-etal-2015}. The prompt $A'$ lines have been observed previously with JWST-MIRI in two disks (d203-506 by \citealt{Zannese-etal-2024}; CX Tau by \citealt{Vlasblom-2024-CXTau}). 

We present the $9 - 12$ $\mu$m spectrum of GM Aur and J1615 in Figure \ref{fig:OHlines}. In pink we highlight the expected locations of prompt transitions associated with quantum numbers $N = 35 - 44$ and $N = 34 - 28$ (wavelengths found in \citealt{Zannese-etal-2024} and \citealt{Carr-Najita-2014}). The specific $A'$ and $A''$ transitions are taken from \cite{Vlasblom-2024-CXTau}. As prompt emission is not formed in LTE, we do not use our slab models to fit this emission. To quantify our detections, we calculate the standard deviation of the difference between the continuum level (as calculated in Section \ref{slabmodels}) and the data between 15.90 - 15.94 $\mu$m, where we expect little to no molecular emission. If the potential line peak is above the nearby continuum level plus five times the standard deviation, we consider it a positive detection. If the peak is within 3- and 5-$\sigma$, it is a tentative detection. Anything below 3-$\sigma$ is a non-detection. 

Visually, the OH emission between $9-12$ $\mu$m is brighter in GM Aur than in J1615. All of the lines labeled in Figure \ref{fig:OHlines} are detected in GM Aur. In J1615, $86\%$ are positive detections, $7\%$ are tentative detections, and $7\%$ are non-detections. Importantly, the $A'$ symmetries at $9.568$ and $10.232$ $\mu$m are detected in both disks.


\subsection{Spectral Energy Distribution Modeling} \label{sed}

\begin{figure*}
\includegraphics[width=0.5\textwidth]{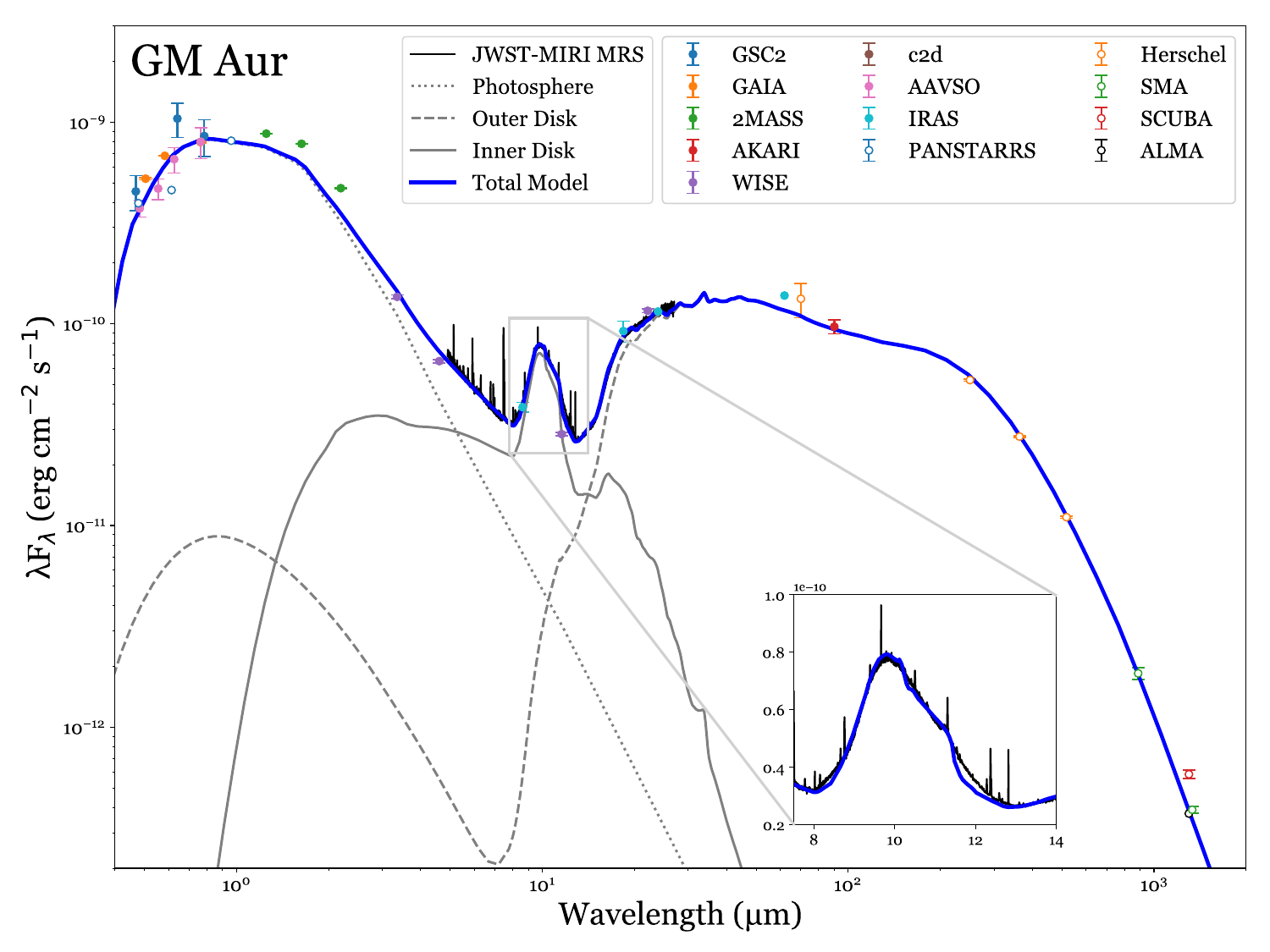}
\includegraphics[width=0.5\textwidth]{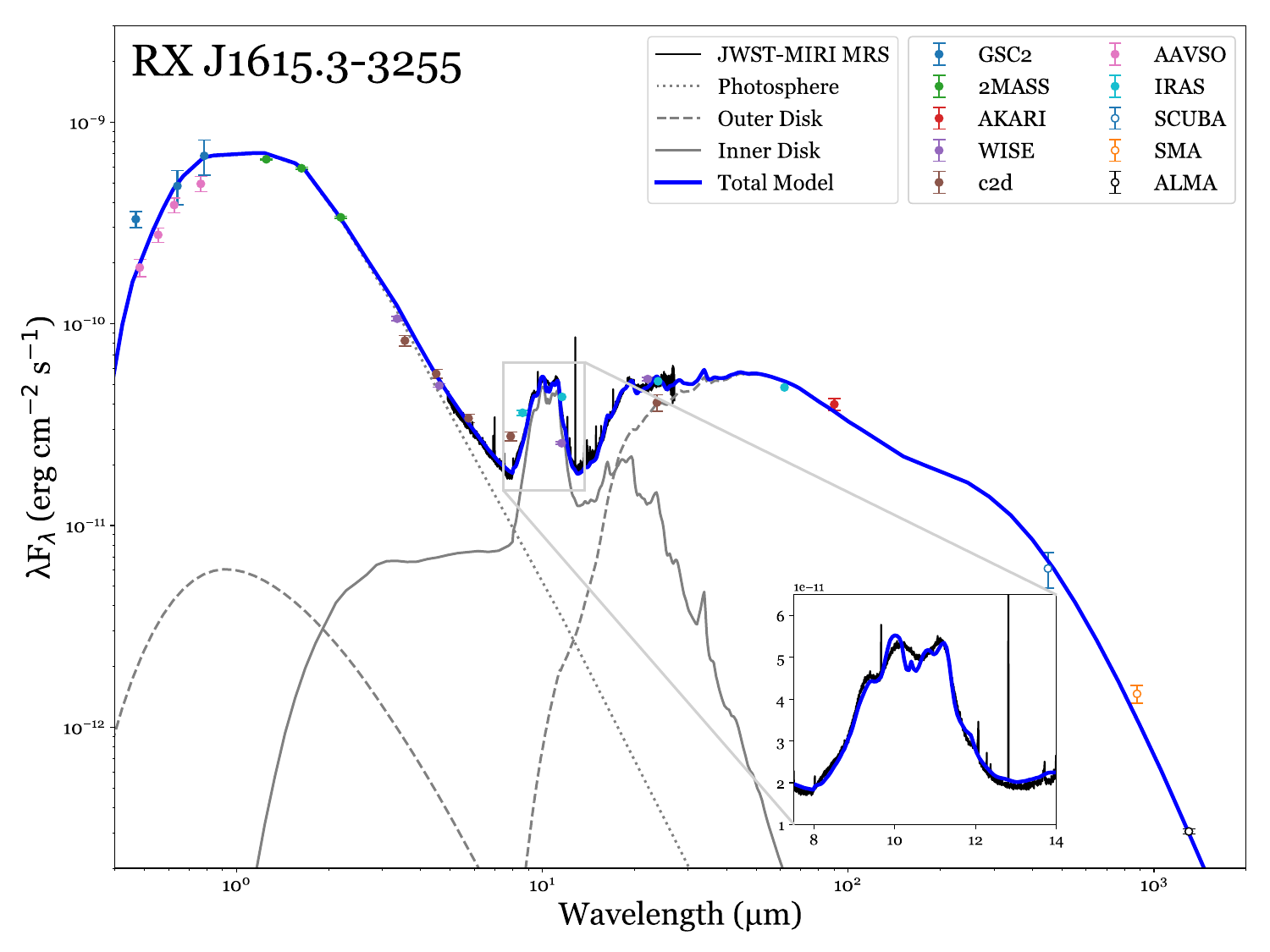}
\caption{SED disk models for GM Aur (left) and J1615 (right). The total model (\textit{blue}) consists of the photosphere (\textit{dotted gray line}), outer disk (\textit{dashed gray lines}), and the optically thin inner dust component (\textit{solid gray line}). Best-fit parameters can be found in Section \ref{sed} and Table \ref{table:bfoptthick}. Photometry is de-reddened using the \cite{Mathis-1999} extinction law with $A_{V} = 0.0$ for J1615 and $A_{V} = 0.6$ for GM Aur \citep[both from][]{Manara-etal-2014}. Labels refer to the observing mission/instrument and their references are included in Table \ref{table:PhotometrySources} in Appendix \ref{photo}.}
\label{fig:SEDs}
\end{figure*}

Another important difference in the JWST-MIRI MRS spectra of GM Aur and J1625 is the appearance of their 10-$\mu$m silicate emission features. Physical differences in the composition of the disk dust may explain these differences. In this section, we present new modeling of the SED of J1615 alongside an updated SED model of GM Aur based on the parameters from \cite{Espaillat-etal-2011}. Our results are shown in Figure \ref{fig:SEDs}. We use the D'Alessio Irradiated Accretion Disk models \citep[DIAD;][]{dAlessio-etal-1998,dAlessio-etal-1999,dAlessio-etal-2001,dAlessio-etal-2005,dAlessio-etal-2006}, which estimates the SED of a disk using an $\alpha$-disk prescription and by self-consistently solving the hydrostatic equilibrium and disk energy transport equations. The input parameters are as follows:
stellar mass ($M_*$), effective temperature ($T_{eff}$), radius ($R_*$),
mass accretion rate ($\dot{M}$),
disk viscosity ($\alpha$),
dust settling ($\epsilon$),
maximum grain size in the disk midplane ($a_{max, mid}$) and disk surface ($a_{max, surf}$),
disk grain size power law $a^{p}$ (for the outer disk, $p = 3.5$),
disk radius ($R_{disk}$),
disk inclination angle ($i$),
dust sublimation temperature ($T_{wall}$), which sets the inner radius of the disk,
and scale height of the inner wall ($z_{wall}$). We direct the reader to \cite{Ribas-etal-2020} for more information on the individual effects of each of these input parameters on the appearance of the SEDs.

The input values used to generate the outer disk component of the SED models in Figure \ref{fig:SEDs} are summarized in Table \ref{table:bfoptthick}. We adopt the same dust-to-gas mass ratio for both targets ($\zeta_{dust} = 0.01$). One can also set the abundances of the silicate dust mixture within the disk. For simplicity, we adopt the same values for both GM Aur and J1615. The mass fractions for amorphous olivine and pyroxene and crystalline forsterite and enstatite for both targets are $f_{oli}= 45\%$, $f_{pyr}= 45\%$, $f_{for}= 5\%$, and $f_{ent}= 5\%$.

\begin{table}[]
\begin{tabular}{lll}
\hline
& GM Aur & RX J1615.3-3255 \\
\hline
\hline
$\alpha$                & 0.0025   & 0.00085       \\
$\epsilon$              & 1.0      & 0.008         \\
$a_{max,mid}$ [$\mu$m]  & 1000     & 1000          \\
$a_{max,surf}$ [$\mu$m] & 3        & 10            \\
$R_{disk}$ [au]         & 150      & 150           \\
$i$ [deg]               & 53$^{a}$ & 47$^{b}$      \\
$T_{wall}$ [K]          & 120      & 110           \\
$z_{wall}$ [au]         & 2.2      & 1.7          \\
$a_{max,wall}$ [$\mu$m] & 3        & 3             \\
$M_{disk}$ [\msun]      & 0.1      & 0.05          \\
\hline
\end{tabular}
\caption{The input parameters used to create the outer disk component of the SED models presented in Figure \ref{fig:SEDs}. The disk inclinations are from (a) \cite{Macias-etal-2018} and (b) \cite{deBoer-etal-2016}.}
\label{table:bfoptthick}
\end{table}

The inner cavity of transitional disks is sometimes populated by optically thin dust, as indicated by the presence of the 10-$\mu$m silicate emission feature \citep{Espaillat-etal-2014}. We follow the work of \cite{Calvet-etal-2002} and generate an optically thin inner dust disk with evenly distributed dust. The parameters of this region include the inner and outer radii,
optical depth ($\tau$),
maximum dust grain size ($a_{max}$; the minimum grain size is held fixed at 0.005~{\micron}),
and the mass fractions of olivine, pyroxene, forsterite, enstatite, and silica.
For more detailed information on the optically thin model, we direct the reader to \cite{Calvet-etal-2002}, \cite{Espaillat-etal-2010}, and \cite{Espaillat-etal-2011}.

For GM Aur, the parameters for the inner disk are taken from \cite{Espaillat-etal-2011} with some modifications to match the JWST-MIRI data. They include an inner radius $R_{in} = 0.17$ au, outer radius $R_{out} =1.0$ au, optical depth $\tau = 0.03$, and maximum dust grain size $a_{max} = 1.0$ $\mu$m. The inner disk of GM Aur is populated by $3 \times 10^{-12}$ \msun\ of dust made up of 32\% organics, 12\% troilite, 55\% silicates, and $<1\%$ forsterite, enstatite, and amorphous carbon. In the case of J1615, its inner disk region spans from $R_{in} = 0.13$ au to $R_{out} = 1.8$ au. Its optical depth is $\tau = 0.04$. The inner dust population has maximum grain size $a_{max} = 10$ $\mu$m, ten times larger than that of GM Aur. Its inner disk dust is comprised of 8.6\% organics, 13\% troilite, 58\% silicates, 8.3\% enstatite, 11.4\% forsterite, and $<1\%$ amorphous carbon.  The total mass of the inner disk dust of J1615 is $2 \times 10^{-11}$ \msun.



\subsection{Atomic Gas Emission Lines} \label{llum}

\begin{figure*}    
    \epsscale{1.2}
    \includegraphics[width = 1.0\textwidth]{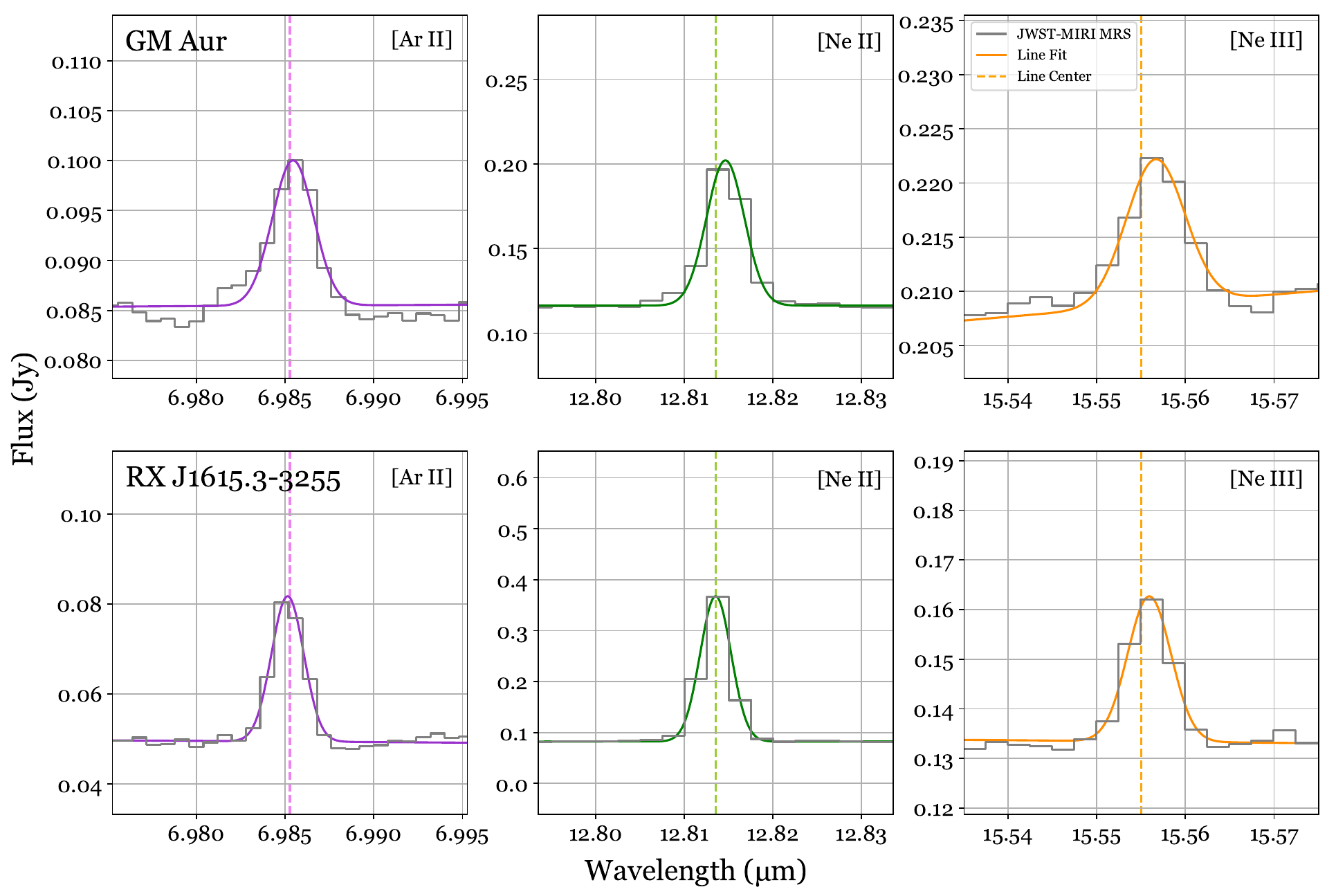}
    \caption{Line fits (\textit{colored lines}) to [\ion{Ar}{2}] 6.99~{\micron}, [\ion{Ne}{2}] 12.81~{\micron}, and [\ion{Ne}{3}] 15.55~{\micron} (\textit{dashed lines}) in the JWST-MIRI MRS spectra of GM~Aur (\textit{gray, top}) and J1615 (\textit{gray, bottom}).}
    \label{fig:atomicfits}
\end{figure*}

The [\ion{Ar}{2}] 6.99~{\micron}, [\ion{Ne}{2}] 12.81~{\micron}, and [\ion{Ne}{3}] 15.55~{\micron} lines are identified in Figure~\ref{fig:fullspec}. Following the same procedure as \cite{Espaillat-etal-2023}, we fit the atomic argon and neon lines using a composite, SciPy \texttt{curve\_fit}-based model of a Gaussian line fit and a linear underlying function for the continuum, which are applied within $\pm 700$ km $s^{-1}$ of the line center. The respective line fluxes for each feature are calculated using the width and amplitude of the Gaussians. Their associated errors are found by propagating the error of the Gaussian fit performed by \texttt{curve\_fit}. The results of this fitting are shown in Figure~\ref{fig:atomicfits}, and the line fluxes are reported in Table~\ref{table:atomicfits}. Our [\ion{Ar}{2}], [\ion{Ne}{2}], and [\ion{Ne}{3}] values are consistent with previous measurements made by \cite{Szulagyi-etal-2012} using \textit{Spitzer} data. 

\begin{table*}[]
\hspace{-3em}
\begin{tabular}{cccccc}
\hline
Target    & [\ion{Ar}{2}] 6.99~{\micron} & [\ion{Ne}{2}] 12.81~{\micron} & [\ion{Ne}{3}] 15.55 $\mu$m & [\ion{Ne}{3}]/[\ion{Ne}{2}] & [\ion{Ne}{2}]/[\ion{Ar}{2}] \\ 
 & (10$^{-15}$ erg cm$^{-2}$ s$^{-1}$) & (10$^{-15}$ erg cm$^{-2}$ s$^{-1}$) & (10$^{-15}$ erg cm$^{-2}$ s$^{-1}$) & Ratio & Ratio \\
\hline
\hline

GM Aur & 2.6 $\pm$ 0.6 & 8.4 $\pm$ 0.6 & 1.4 $\pm$ 0.4 & 0.16 $\pm$ 0.05 & 3.2 $\pm$ 0.7 \\

RX J1615.3-3255 & 4.5 $\pm$ 0.6 & 22.8 $\pm$ 0.8 & 2.2 $\pm$ 0.4 & 0.10 $\pm$ 0.02 & 5.1 $\pm$ 0.7 \\

\hline
\end{tabular}
\caption{Atomic line fluxes for GM Aur and J1615.}
\label{table:atomicfits}
\end{table*}

Atomic emission may be produced by jets \citep[e.g.,][]{Baldovin-Saavedra-etal-2012}. We do not find evidence of a jet in the GM Aur system based on the MIRI Integral Field Unit (IFU) images. This is consistent with the analysis of GM Aur's \textit{Spitzer} data performed by \cite{Najita-etal-2009}, who found that GM Aur's [\ion{Ne}{2}] line is centered at the stellar velocity. The same holds for J1615; there is no extended emission in the MIRI-IFU images, consistent with \cite{Sacco-etal-2012}, who identify slow disk winds as the source of the [\ion{Ne}{2}] blueshift ($-7.5 \pm 2.8$ km s$^{-1}$) in J1615. 

Our measured [\ion{Ne}{3}]/[\ion{Ne}{2}] and [\ion{Ne}{2}]/[\ion{Ar}{2}] flux ratios for GM Aur are $0.16 \pm 0.05$ and $3.2 \pm 0.7$, respectively. The [\ion{Ne}{3}]/[\ion{Ne}{2}] and [\ion{Ne}{2}]/[\ion{Ar}{2}] flux ratios for J1615 are $0.10 \pm 0.02$ and $5.1 \pm 0.7$, respectively. 
These ratios are consistent with X-ray photoevaporation \citep{Glassgold-etal-2007, Hollenbach-etal-2009}. 


\subsection{Gas Accretion Rates} \label{acc}

Accretion rates for GM Aur and J1615 are measured using a magnetospheric accretion flow model \citep{hartmann94,muzerolle98b,muzerolle01} fit to the continuum-normalized Chiron \halpha\ profiles taken contemporaneously with the JWST observations. The model assumes an axisymmetric dipole field that channels gas from the inner disk at an infall radius \ri\ with a width of \rw\ at the disk midplane. The flow has a maximum temperature \tmax, an accretion rate \mdot, and is viewed at an angle $i$ relative to the pole. Here, we calculate a grid of 302,400 models for J1615, using the stellar parameters in Table~\ref{table:stellar-parameters}. We use the existing grid of 525,097 models presented in \cite{Wendeborn-etal-2024c} for GM~Aur. We use the \chisq\ metric to calculate the goodness-of-fit of each model to the \halpha\ profiles, masking out any regions with narrow absorption that cannot be explained by the flow model. Then, we weight each model by $\exp(-\chi^{2}/2)$ to measure the median and standard deviation of each parameter for each observation. We confirm that the best-fit parameter distributions do not fall at the edges of the grid. Finally, we calculate a weighted average over the five profiles to determine a ``typical'' configuration of the accretion system. 

\begin{figure*}
    \centering
    \includegraphics[width=0.6\textwidth]{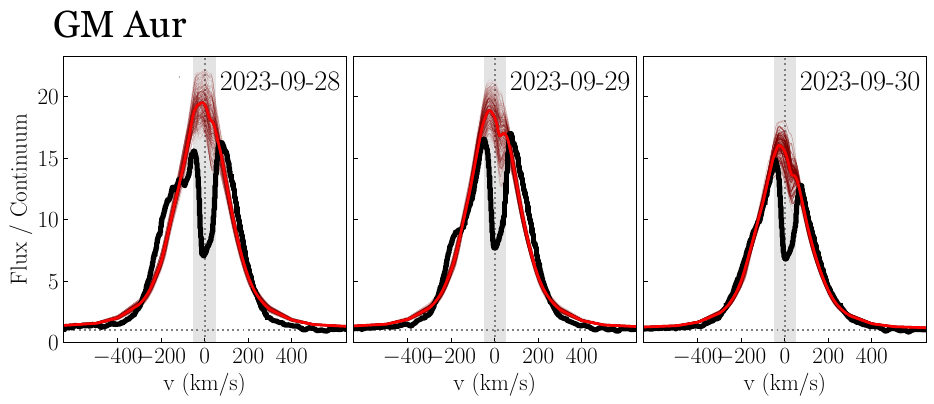}
    
    \includegraphics[width=\textwidth]{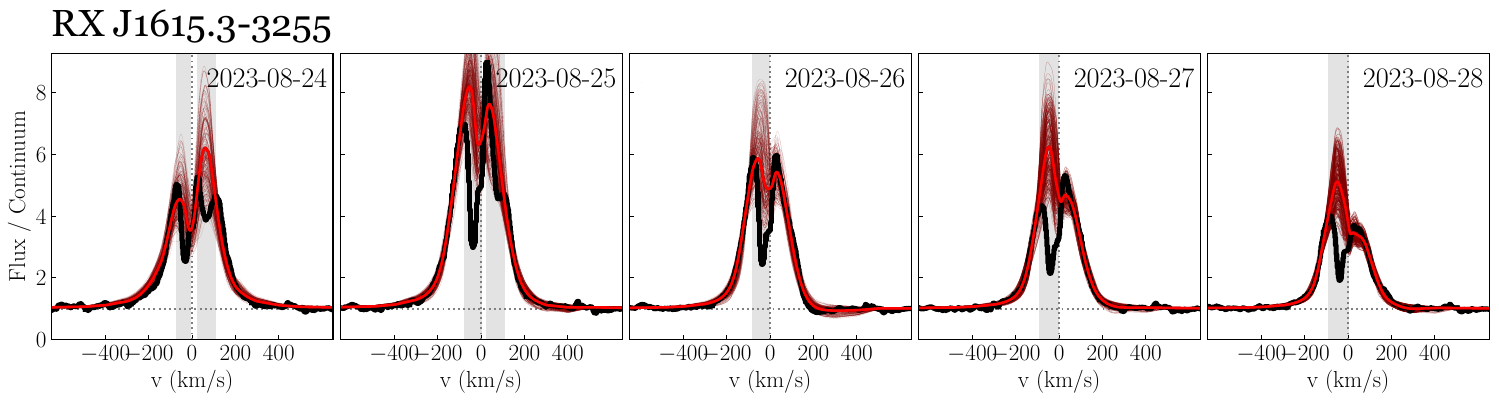}
    \caption{Accretion flow model fits (\textit{red}) to the Chiron \halpha\ observations (\textit{black}) surrounding the JWST observations of J1615 (top row) and GM~Aur (bottom row). The low-opacity maroon lines show the 200 best-fit models from which the parameter medians and standard deviations are determined (presented in Table~\ref{table:accretion-parameters}). Absorption that is not taken into account by the flow model is masked out, indicated by the gray shaded regions.}
    \label{fig:flow_fits}
\end{figure*}

We report the results in Table~\ref{table:accretion-parameters} and show the fits in Figure~\ref{fig:flow_fits}.
We find that the accretion luminosities that we measure from our accretion flow models are consistent with those derived using JWST mid-infrared \ion{H}{1} transition lines (10-7), (7-6), and (8-7) following \cite{Tofflemire-etal-2025}.

\begin{table*}[]
\begin{tabular}{llccccc}
\hline
Target & Observation Time & \mdot\ [$10^{-9}$ \msunyr] & \ri\ [\rstar] & \rw\ [\rstar] & \tmax\ [K] & $i$ [$^\circ$] \\
\hline
\hline
GM Aur & 2023-09-28T08:55:49.1\tablenotemark{$\star$} & $32.00\pm12.30$ & $4.43\pm0.26$ & $0.50\pm0.17$ & $8190\pm384$ & $60\pm3$ \\
 & 2023-09-29T08:49:16.7 & $27.40\pm13.70$ & $3.93\pm0.23$ & $0.36\pm0.12$ & $8438\pm731$ & $52\pm4$ \\
 & 2023-09-30T08:42:06.7 & $15.60\pm11.60$ & $3.49\pm0.29$ & $0.56\pm0.23$ & $9458\pm1083$ & $57\pm4$ \\
 \cmidrule(lr){2-7}
 & Weighted average & $24.40\pm7.19$ & $3.98\pm0.15$ & $0.43\pm0.09$ & $8353\pm324$ & $58\pm2$ \\
\midrule
RX J1615.3-3255 & 2023-08-24T00:17:11.8 & $2.81\pm1.93$ & $5.73\pm1.34$ & $0.79\pm0.57$ & $9011\pm785$ & $82\pm16$ \\
 & 2023-08-25T00:40:10.9\tablenotemark{$\star$} & $3.28\pm2.32$ & $4.55\pm1.23$ & $0.93\pm0.56$ & $9200\pm951$ & $62\pm11$ \\
 & 2023-08-26T01:11:17.2 & $2.61\pm1.84$ & $4.21\pm0.84$ & $1.06\pm0.54$ & $9452\pm806$ & $54\pm12$ \\
 & 2023-08-27T00:56:32.5 & $2.17\pm1.35$ & $3.77\pm2.29$ & $0.89\pm0.51$ & $9995\pm996$ & $41\pm23$ \\
 & 2023-08-28T00:51:28.7 & $2.48\pm1.31$ & $2.27\pm1.39$ & $0.76\pm0.42$ & $10505\pm903$ & $28\pm16$ \\
\cmidrule(lr){2-7}
 & Weighted average & $2.54\pm0.73$ & $4.21\pm0.55$ & $0.87\pm0.23$ & $9582\pm392$ & $55\pm6$ \\
\bottomrule
\end{tabular}
\caption{Best-fit flow model parameters and standard deviations for GM~Aur and J1615. We include results for individual epochs of observation as well as a weighted-average of the nightly results. The corresponding fits are shown in Figure~\ref{fig:flow_fits}.\\
A star denotes the spectrum closest in time to the JWST observation.
}
\label{table:accretion-parameters}
\end{table*}


\subsection{Dust Continuum Variability} \label{cont}

\begin{figure*}    
\epsscale{1.2}
\includegraphics[width = 1.0\textwidth]{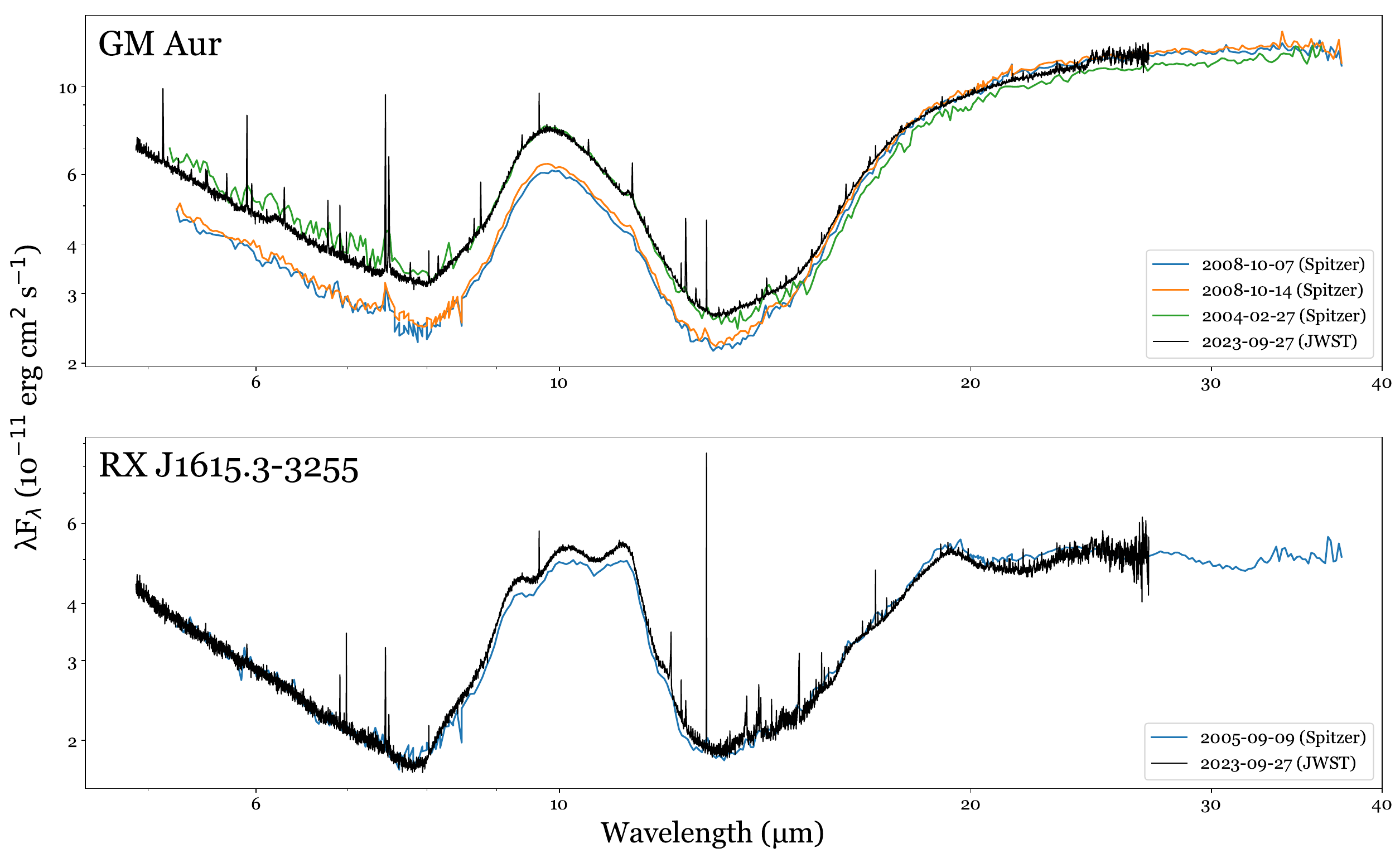}
\caption{JWST-MIRI MRS spectra (\textit{black lines}) of GM Aur and J1615 plotted alongside their respective archival \textit{Spitzer} spectra (\textit{colored lines}). GM Aur shows continuum variability in the shorter wavelengths, while J1615's continuum level is consistent between \textit{Spitzer} and JWST observations.}
\label{fig:variability}
\end{figure*}

In Figure \ref{fig:variability} we compare the JWST spectra of GM Aur and J1615 to low-resolution \textit{Spitzer} spectra taken from the Combined Atlas of Sources with \textit{Spitzer} IRS Spectra \citep[CASSIS,][]{Lebouteiller-etal-2015}. The J1615 \textit{Spitzer} spectrum is from GO Program 179 \citep[PI: Evans;][]{Merin-etal-2010}.  For GM Aur, the two 2008 \textit{Spitzer} spectra were taken by GO Program 50403 \citep[PI: Calvet;][]{Espaillat-etal-2011}, and the 2004 spectrum was observed by GTO Program 2 \citep[PI: Houck;][]{Furlan2011}.

GM Aur has continuum variability at wavelengths below $\sim$17 $\mu$m, while there is no significant variability in the MIR continuum emission of J1615. Interestingly, the level of the MIR emission seen in the JWST spectrum of GM Aur taken on 2023-09-27 is within the range seen previously with \textit{Spitzer} and is roughly consistent with the 2004 spectrum. The JWST spectrum taken as part of JWST General Observers program 2025 (P.I. Öberg) on 2023-10-14 and presented by \cite{Romero-Mirza-etal-2025} is also within this range. This may point to a recurring phenomenon and possibly a minimum and maximum MIR continuum flux level. MIR variability has been previously explained with changes in the amount of dust in the inner disk or geometrical changes in the inner disk \cite{Espaillat-etal-2011, Espaillat-etal-2024}.  More data is necessary to discern the underlying cause of the variability seen in GM~Aur.


\section{Discussion} \label{disc}
As explained in Section \ref{intro}, there are a growing number of examples of carbon-rich full disks around T Tauri stars. It is less known, however, how population trends change if we narrow our focus to \textit{transitional} protoplanetary disks. To date, out of the 213 candidate transitional disks identified by \cite{vanderMarel-etal-2016}, only 38 have been observed with JWST-MIRI, and only a subset of those have been published. The spectrum of J1615 presented in this work is a new example of a carbon-rich transitional disk. The molecular emission from J1615 is particularly noteworthy for two key reasons. First, it has a high carbon content, with a column density ratio of $N_{CO_{2}}/N_{H_2O} = 2.15$--orders of magnitude higher than the median value of $5 \times 10^{-4}$ reported by \citet{Salyk-etal-2011} for 48 T Tauri stars. Second, J1615 displays strong emission lines from carbon-bearing species, including HCN and C$_2$H$_2$. Notably, strong C$_2$H$_2$ is typically observed in disks around low-mass stars ($< 0.3$ \msun) \citep{Pascucci-etal-2009, Grant-etal-2025}. 

We chose to present J1615 alongside GM Aur to highlight the diversity of chemical inventories in transitional disks. Both objects are young (1--4 Myr old), late K-type stars \citep[][]{Wahhaj-etal-2010, Koerner-etal-1993, Beckwith-etal-1990} with similar stellar masses \citep[1.1–1.4 \msun;][]{Manara-etal-2014} and X-ray luminosities \citep[$\sim$1$\times$10$^{30}$ erg s$^{-1}$;][]{Espaillat-etal-2021,Krautter-etal-1997}. Both are surrounded by transitional disks with $\sim30-40$ au cavities \citep{Manara-etal-2014, Andrews-etal-2011} that contain some small dust grains, as evidenced by strong 10 {\micron} silicate emission features and some excess emission above the photosphere at shorter MIR wavelengths (Figure \ref{fig:SEDs}). Even with these physical similarities, their MIR spectra are quite different, with GM Aur mostly devoid of molecular species detected in J1615.

Below we explore the possible mechanisms behind J1615’s carbon-rich chemistry, focusing on the two most noticeable differences between GM Aur and J1615, namely their accretion rates and inner disk dust properties. We then compare these two objects to other transitional disks observed with JWST in order to assess whether these differences reflect broader trends among transitional disks.

\subsection{Accretion Rate} \label{acc_disc}

J1615’s accretion rate is about ten times lower than GM Aur’s (Table \ref{table:stellar-parameters}), consistent with previous measurements \citep{Manara-etal-2014, Wendeborn-etal-2024a, Wendeborn-etal-2024b}. There is growing support that low accretion may be tied to the detection of carbon-bearing molecular species. As noted in Section \ref{intro}, C$_2$H$_2$ is more commonly detected in disks around low-mass stars, which typically have lower accretion rates \citep{Manara-etal-2016, Hartmann-2016, Alcala-etal-2017}. In solar-mass stars, lower accretion rates have been proposed as a mechanism for observing carbon-rich emission \citep{Colmenares-etal-2024}. \cite{JDISC-2025} find strong correlations between the emission line luminosities for HCN, C$_2$H$_2$, and CO$_2$ and the mass accretion rate of the central star in a sample of 31 T Tauri stars. \cite{Grant-etal-2025} also find a correlation between the C$_2$H$_2$ flux and accretion rate in a sample of 34 very low-mass star and T Tauri star disks.
Below, we examine how a lower accretion rate in J1615 could enhance carbon-bearing molecular emission by influencing disk material transport and/or affecting the UV radiation field.

\subsubsection{Vertical transport} 

Hydrocarbons such as C$_2$H$_2$ form near the disk midplane, deeper in the disk than other species such as H$_2$O and CO$_2$ \citep{Woitke-etal-2018b}, but can be obscured by optically thick dust unless transported to the upper disk layers, a process set by the vertical mixing timescale. The visibility of carbon-bearing species in a disk depends on a close relationship between the vertical mixing timescale and the location of the soot line, where carbonaceous materials sublimate \citep{Kress-etal-2010, Li-etal-2021}. Only at the soot line will carbon in the disk become available for the creation of carbon-bearing species. Typically, the vertical mixing timescale scales with
\begin{equation}
    \tau \propto \frac{h}{c_s \alpha}
\end{equation}
where $h$ is the scale height, $c_s$ is the sound speed, and $\alpha$ is the viscosity of the disk. For a disk with a lower accretion rate, $\alpha$ is typically smaller (consistent with GM Aur and J1615 in Table \ref{table:bfoptthick}), and therefore $\tau$ should take longer, impacting the visibility of sublimated carbon. 

If we replace the scale height with $h = c_s / \Omega$, for a fixed viscosity the vertical mixing timescale instead scales with:
\begin{equation}
    \tau \propto R^{3/2} M^{-1/2}.
\end{equation}
At the same distance $R$ from the central star, the disk around a solar-mass star (1 \msun) would have a timescale around three times faster than a low-mass star (0.1 \msun). However, carbonaceous materials only sublimate at the soot line, so the visibility of any carbon-bearing species is only relevant within that line. As it turns out, the soot line is ten times closer to the low-mass star than it is to the solar mass star. Thus, at the soot line, the low-mass star has a much shorter vertical mixing timescale than the solar-mass star.

With the higher accretion rates expected of solar-mass stars \citep{Hartmann-2016}, vertical mixing is generally too slow to expose midplane C$_2$H$_2$ before it is transported radially and accreted onto the star. An exception is DoAr 33, a 1.1 \msun\ star with a Class II disk, which exhibits strong C$_2$H$_2$ and hydrocarbon emission \citep{Colmenares-etal-2024}, attributed to its unusually low accretion rate (2.5$\times$10$^{-10}$ \msunyr; \citealt{Cieza-etal-2010}). \citet{Colmenares-etal-2024} describe this as a “burn and linger” scenario, where midplane material sublimates (“burns”) as it rises and remains suspended (“lingers”) due to slow accretion.  

J1615’s low $\dot{M}$ is consistent with this scenario. However, both J1615 and GM~Aur possess large optically thin cavities, suggesting that if carbon-bearing material is present in the inner disk, it could potentially be detected regardless of its vertical height.

\subsubsection{FUV photodissociation} \label{FUV}

FUV photons can photodissociate molecules, and the FUV emission of T Tauri stars is known to correlate with accretion rate \citep[e.g.,][]{Calvet-etal-2004, Wendeborn-etal-2024a, Pittman-etal-2025}. \citet{vanDischoeck-etal-2008} note that molecules such as H$_2$O, C$_2$H$_2$, HCN, and CO$_2$ have relatively high collisional cross sections in Ly$\alpha$ radiation fields. In particular, C$_2$H$_2$ exhibits a large cross section at Ly$\alpha$ \citep[see Table 1 of][]{Heays-etal-2017}, even exceeding that of H$_2$O. Consequently, if H$_2$O is photodissociated, C$_2$H$_2$ is likely to be dissociated as well.

GM Aur’s higher accretion rate implies a stronger FUV radiation field compared to J1615, which could photodissociate C$_2$H$_2$ into C$_2$H and H \citep{Heays-etal-2017} and explain its non-detection. Supporting this, GM Aur’s mid-infrared spectrum shows emission from very high-temperature ($>1000$~K) OH, which is indicative of OH production via H$_2$O dissociation driven by Ly$\alpha$ radiation known as ``prompt emission" \citep{Tappe-etal-2008, Carr-Najita-2014, Tabone-etal-2021, Tabone-etal-2024, Vlasblom-2024-CXTau}. As described in Section \ref{oh}, we observe prompt emission of OH in both GM Aur and J1615, but the lines found in GM Aur are visually brighter than those in J1615. J1615 exhibits weaker OH emission, consistent with a lower FUV flux.

Another hint to GM Aur and J1615's relative FUV radiation levels is the detection of the rovibrational band of CH$_3^+$ at 7.15 $\mu$m in the disk of GM Aur (see Figure \ref{fig:6-7um_slabs}). CH$_3^+$ has been observed in the disks of TW Hya \citep{Henning-etal-2024} and d203-506 \citep{Zannese-etal-2024}. The latter credit the efficient production of CH$_3^+$ to the strong FUV-irradiated environment around d203-506 from nearby bright stars. In TW Hya, a transitional disk, thermochemical models point to the CH$_3^+$ emitting from the inner cavity wall where stellar radiation is most intense \citep{Henning-etal-2024}. Thus, GM Aur's detection and J1615's non-detection of CH$_3^+$ may be another sign of the increased FUV levels in GM Aur's disk environment. 

Our detections of OH prompt emission and CH$_3^+$ in the disk of GM Aur are consistent with the findings of \cite{Romero-Mirza-etal-2025}, who find that these species may emit from the edge of a 0.2 au dust cavity. This is also consistent with the inner disk models presented in Section \ref{sed}, which best reproduce GM Aur's SED with an inner radius of $R_{in} = 0.17$ au. \cite{Romero-Mirza-etal-2025} propose that GM Aur is depleted in carbon and water because FUV light from the central star photodissociates the inner disk, leaving bright prompt OH emission, excited H$_2$, \ion{H}{1}, and molecular ions CH$_3^+$ and/or HCO$^+$.

In summary, the stronger FUV field in GM Aur likely promotes photodissociation of carbon-bearing molecules, whereas J1615’s lower accretion rate and diminished UV radiation allow these molecules to survive and be observed through its optically thin inner cavity.

\subsection{Inner disk dust properties}

GM Aur and J1615 also differ in their inner dust disk composition.  A striking difference between the spectra of GM Aur and J1615 is the appearance of their 10-$\mu$m silicate emission features which indicate that J1615 has larger grains and significantly more crystalline silicates than GM Aur.  Additionally, ALMA observations reveal that,  while both GM Aur and J1615 have dust in the inner disk, J1615's inner disk is much brighter \citep{Sierra-etal-2024, Francis-vanderMarel-2020}. Below, we explore how differences in dust grain properties or the inner disk structure may influence the observed carbon-bearing molecular emission.

\subsubsection{Dust opacities} \label{dustopa}

As discussed in Section~\ref{sed}, J1615’s 10-$\mu$m silicate emission feature is best modeled with dust grains approximately ten times larger and with a significantly higher fraction of crystalline material compared to GM Aur. This provides evidence for more advanced dust processing in the disk around J1615.

In their study of disks around very low-mass stars (VLMS) and brown dwarfs, \citet{Arabhavi-etal-2025} found that the mid-infrared molecular emission depends sensitively on the location of the $\tau_{dust} = 1$ layer. As disks evolve, grain growth reduces dust opacity and shifts this $\tau_{\rm dust} = 1$ layer deeper into the disk. According to the disk structure models of \citet{Woitke-etal-2018b}, the majority of emissive water resides near the disk surface, followed by CO$_2$ and then C$_2$H$_2$ at progressively deeper layers. The position of the $\tau_{\rm dust} = 1$ layer thus determines which molecular species produce the brightest mid-infrared features. For J1615, this layer may lie between the CO$_2$ and C$_2$H$_2$ emitting regions, consistent with its relatively strong CO$_2$ emission compared to H$_2$O and the absence of pseudo-continuum emission from C$_2$H$_2$ \citep[as seen in VLMS ISO-ChaI 147;][]{Arabhavi-etal-2024}. In contrast, GM Aur exhibits less grain growth, resulting in larger dust opacity, a vertically higher $\tau_{\rm dust} =1$ layer, and correspondingly dimmer H$_2$O, CO$_2$, and C$_2$H$_2$ emission.

The high crystalline fraction of J1615's inner disk ($\sim 20\%$, see Section \ref{sed}) also has important implications on the opacity. While amorphous silicates and crystalline forsterite and enstatite absorb similarly near 10 $\mu$m, the crystalline species become more opaque than the amorphous species at UV wavelengths \citep{Dorschner-etal-1995, Chihara-etal-2002, Sogawa-etal-2006}. Thus, the crystalline silicates in J1615's inner disk would absorb a higher fraction of the stellar FUV radiation, which emits from J1615's accretion shock at an already smaller amount compared to GM Aur due to J1615's lower accretion rate. This would protect the water and carbon-bearing molecules in J1615's disk from photodissociation, enhancing the effects described in Section \ref{FUV}. 

In addition to larger and more crystalline dust grains in the inner disk of J1615, its outer disk is best modeled with a smaller $\epsilon$ than GM Aur (see Table \ref{table:bfoptthick}), which indicates more advanced dust settling. As settling increases, \cite{McClure-etal-2016} find that the hot upper layer and cold midplane of the disk shrink, while the middle ($\sim 30-300$ K) layer expands (see their Figure 5). The range of temperatures in this layer happens to match the best-fit temperatures of the carbon-bearing species in J1615 (see Table \ref{table:bestfit}). Combined with the opacity of crystalline silicates, this thick carbon-bearing layer may be more protected from FUV radiation in J1615 than in GM Aur.

In summary, the larger, more processed dust grains in J1615 may reduce inner disk MIR opacities, facilitating the detection of carbon-bearing molecules from the midplane. At the same time, J1615's large crystalline fraction would protect its potentially vertically thicker layer of carbon-bearing species from FUV photodissociation. However, since both disks feature optically thin cavities extending tens of au, it is unclear if this difference in dust opacity would have a significant observational impact.

\subsubsection{Disk structure} \label{drift_disc}

\begin{figure*}
    \includegraphics[width = 1.0\textwidth]{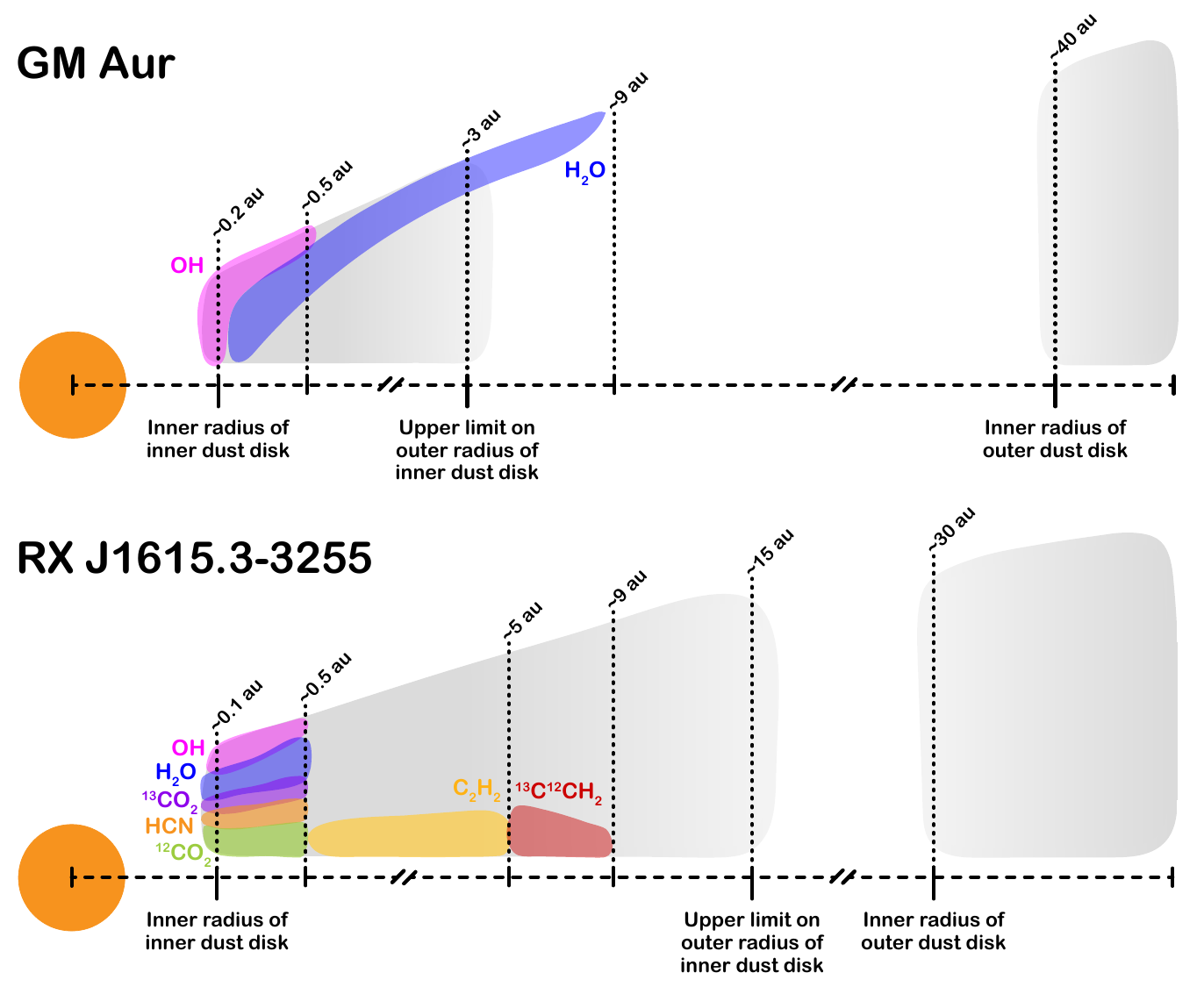}
    \caption{Schematics of the disks of GM Aur (top) and J1615 (bottom). In the lower panel, the order of the species from disk surface to midplane follows with increasing best-fit temperatures in Table \ref{table:bestfit}. The right-hand edge of each colored region for each species correlates with the approximate location of the best-fit emitting radius described in Section \ref{slabmodels} and Table \ref{table:bestfit}. Note that the emitting radius may correspond to a full disk of emission with $R$ or a ring with an area equal to $\pi R^2$. The gray shaded areas represent disk radii where dust is present. The base of the disk cross-section extends only to the midplane. For GM Aur, the inner dust disk has been observed to extend up to 3.2 au \citep{Francis-vanderMarel-2020} and up to 15 au in J1615 \citep{Sierra-etal-2024}. The inner radius of the inner dust disks in both targets corresponds to the inner radius modeled in Section \ref{sed}.}
    \label{fig:schematic}
\end{figure*}

There are several ways that pebble drift and dust traps may affect carbon emission in disks \citep[e.g.,][]{Sellek-etal-2025, Vlasblom-etal-2024, Bosman-etal-2017, Banzatti-etal-2020}. Here we focus on the inner disk since the outer disks of GM Aur and J1615 are fairly similar \citep[i.e., large cavities and multiple rings;][]{deBoer-etal-2016, Sierra-etal-2024, Huang-etal-2020}. To better illustrate our arguments, we include simplified schematics of the GM Aur and J1615 disks in Figure \ref{fig:schematic}. The approximate emitting radii derived from the slab models in Section \ref{slabmodels} are represented with colored shapes labeled by species. The dust morphology of each target as probed by ALMA \citep{Sierra-etal-2024, Macias-etal-2018, Huang-etal-2020, Francis-vanderMarel-2020} is also illustrated. We note that the ALMA observations are tracers of both thermal dust \textit{and} free-free emission from jets and/or winds emanating form the central star \citep{Rota-etal-2024, Rota-etal-2025}. While free-free emission may contribute to GM Aur's inner disk emission, the spectral index of its inner disk ($\alpha \sim2.7$) suggests that it must be supplemented by a significant dust contribution \citep{Huang-etal-2020}. For J1615, there is no constraint in the literature on the spectral index of the inner disk, but the brightness of its inner region also suggests a significant dust contribution \citep{Sierra-etal-2024}.

As mentioned in Section \ref{h2o} and \ref{disc}, both GM Aur and J1615 are water-poor disks compared to other full and transitional disks. \cite{Temmink-etal-2025} suggest that water depletion could be a result of a small inner cavity depleting the warm-temperature reservoirs. This suggestion was first made by \cite{Grant-etal-2023} as a possible explanation for the CO$_2$-rich spectrum of GW Lup, and was later tested by \cite{Vlasblom-etal-2024} who modeled the effects of cavities of various radii on the brightness of CO$_2$ vs. H$_2$O MIR features. In smooth disks, both CO$_2$ and H$_2$O ice can drift unobstructed across their respective snowlines, at which point they will sublimate and emit, resulting in a H$_2$O-bright spectrum. When a cavity that terminates between the H$_2$O and CO$_2$ snowlines is present in the disk, however, it impedes the drift of water ice across its snowline, prevents its sublimation, and decreases hot water emission in the MIR. They also found that the snowlines of disks with inner cavities are pushed out to larger radii, which could contribute to the surviving reservoirs of cold water in ``H$_2$O-poor" disks. Perhaps, within the compact inner disk of J1615, there exists a small inner cavity between the water and CO$_2$ snowlines, generating the bright-CO$_2$ and dim-H$_2$O we see in its MIR spectrum. 

Another piece of supporting evidence to the existence of a drift impediment near the water snowline is the presence of HCO$^+$ emission in J1615. HCO$^+$ is destroyed by gaseous water \citep{Phillips-etal-1992, Bergin-etal-1998}:
\ce{HCO+ + H2O \rightarrow CO + H3O+}
In other words, as more water enters the solid phase (such as at the snowline; \citealt{Leemker-etal-2021}), HCO$^+$ becomes more abundant. HCO$^+$ emission in J1615 and not in GM Aur is suggestive of higher abundances of water in J1615 than in GM Aur, or perhaps a ``pile-up" or water ices stuck outside the snowline and unable to sublimate due to an inner cavity \citep{Vlasblom-etal-2024}. 

Another scenario comes from \cite{Sellek-etal-2025b}, who find that carbon can be enhanced in a warm dust trap at less than 5 au.  In short, a dust trap between the CH$_4$ and CH$_3$OH snowlines would allow CH$_3$OH photodissociation to occur and provide carbon in the inner disk while also blocking O-bearing species such as H$_2$O and CO$_2$ from sublimating and emitting.  These so-called ``warm" dust traps have been suggested to explain extreme C/O ratios such as in DoAr 33 \citep{Colmenares-etal-2024}, but there is no detection of CH$_4$ in J1615.

In the case of J1615, ALMA observations have revealed a compact inner disk as traced by emission at $\sim0.8-1.1$ mm, indicating large grains in the inner disk. \cite{Sierra-etal-2024} present a brightness profile of the disk of J1615, which indicates emission out to at most $\sim15$ au. In terms of J1615's SED, however, there is no substantial excess emission above the photosphere, which suggests that the compact inner disk does not extend all the way to the dust sublimation radius. With an optically thin inner disk at the dust sublimation radius and an outer radius of $R_{out} = 1.8$ au, our inner disk model may overlap with the compact inner disk detected by ALMA, but resolution limitations cannot resolve the structure of J1615 down to au scales. Nonetheless, J1615's carbon-richness may hint at smaller substructures around the water, CO$_2$, and CH$_4$ snowlines.

In summary, J1615's inner disk may host further substructures, such as a cavity between the H$_2$O and CO$_2$ or CH$_3$OH and CH$_4$ snowlines, and help explain J1615’s carbon-rich chemistry.

\subsection{Comparison to other transitional disks}

In this section, we contextualize the molecular inventories of GM Aur and J1615 against published JWST-MIRI observations of other transitional disks: MY Lup \citep{Salyk-etal-2025}, SY Cha \citep{Schwarz-etal-2024}, PDS 70 \citep{Perotti-etal-2023}, TW Hya \citep{Henning-etal-2024}, T Cha \citep{Xie-etal-2025, Bajaj-etal-2024}, SZ Cha \citep{Espaillat-etal-2023}, UX Tau A \citep{Espaillat-etal-2024}, RY Lup, AS 205 S, DoAr 25, HP Tau, IRAS 04385+2550, and SR 4 \citep{JDISC-2025}.

Water emission is well-represented in the transitional disks around T Tauri stars. 
Of the spectra available in literature, MY Lup \citep{Salyk-etal-2025, JDISC-2025}, SY Cha \citep{Schwarz-etal-2024}, PDS 70 \citep{Perotti-etal-2023}, TW Hya \citep{Henning-etal-2024}, HP Tau \citep{Romero-Mirza-etal-2024b}, RY Lup, AS 205 S, DoAr 25, HP Tau, and SR 4 \citep{JDISC-2025} all have positive water detections. PDS 70’s brightest water features are found shortward of 15 $\mu$m, and are well fit with a single, warm-temperature (600 K) component, suggesting that there is no additional cool water reservoir in the disk. MY Lup’s water levels are low compared to its carbon-bearing species, similar to J1615. 
TW Hya is similar to GM Aur in regards to its bright prompt OH features and molecular and ionic hydrogen alongside subdued water lines. SY Cha and HP Tau are examples of water-rich transitional disks, with the former having a CO$_2$/H$_2$O column density ratio of $\sim0.06$ in the wavelength range $13.5-15.5$ $\mu$m \citep{Schwarz-etal-2024}, and the latter with a ratio of $\sim0.003$ between $12-16$ $\mu$m \citep{JDISC-2025}. This is much lower than that derived for J1615 in a similar wavelength range ($\sim 2.15$, see Sections \ref{slabmodels} and \ref{disc}). 
T Cha \citep{Xie-etal-2025, Bajaj-etal-2024}, SZ Cha  \citep{Espaillat-etal-2023}, and UX Tau A \citep{Espaillat-etal-2024} do not have any reported water detections. Interestingly, all show evidence of variable stellar emission impinging on their inner disks, whether it be from a sudden stellar outburst (T Cha), a changing stellar wind (SZ Cha), or a misaligned inner disk (UX Tau A).

While JWST has observed several transitional disks, only one exhibits strong carbon emission comparable to J1615: MY Lup. MY Lup has reported accretion rates between $10^{-8} - 2 \times 10^{-10}$ \msunyr \citep{Alcala-etal-2019, Alcala-etal-2017} and a disk with bright CO$_2$, HCN, and their isotopologues. Its carbon-rich chemistry is believed to be a result of inner-disk clearing of water molecules and its nearly edge-on inclination, which exposes larger columns of carbon-bearing species \citep{Salyk-etal-2025}. The disks of DoAr 25, IRAS 04385+2550, and HP Tau ($\log{(\dot{M})} = -8.9$, $-8.1$, and $-10.3$ \msunyr, respectively; \citealt{Manara-etal-2023}) have positive detections of both CO$_2$ and C$_2$H$_2$ alongside comparably bright water lines \citep{JDISC-2025}. 
Other disks show detections of only CO$_2$ or C$_2$H$_2$.  These include
AS 205 S \citep{JDISC-2025},
SR 4 \citep[accreting at $\log{(\dot{M})} = -6.9$ \msunyr;][]{Manara-etal-2023},
SY Cha \citep[accreting at $\log{(\dot{M})} = -9.18$ \msunyr;][]{Manara-etal-2023},
PDS 70 \citep[accreting at $\log{(\dot{M})} \approx -10$ \msunyr;][]{Thanathibodee-etal-2020}, and
TW Hya \citep[accreting at $\log{(\dot{M})} \approx -8.7$ \msunyr\ ;][]{Manara-etal-2014, Wendeborn-etal-2024a}.
Other transitional disks observed by JWST lack CO$_2$ and C$_2$H$_2$ detections based on visual inspection of their published spectra. These include T Cha \citep{Xie-etal-2025, Bajaj-etal-2024}, SZ Cha \citep{Espaillat-etal-2023}, UX Tau A \citep{Espaillat-etal-2024}, and RY Lup \citep{JDISC-2025}, all of which have accretion rates of order 10$^{-8}$ \msunyr \citep{Schisano-etal-2009, Cahill-etal-2019, Manara-etal-2023, Espaillat-etal-2024, Gahm-etal-1993}.

Considering accretion alone, SY Cha and PDS 70 both have low accretion rates but lack strong carbon emission. Conversely, SR 4 is a fast accretor with visible C$_2$H$_2$, suggesting that accretion rate by itself does not determine carbon abundance in transitional disks. We also inspected the 10-$\mu$m silicate emission features for the disks above in their published spectra 
(SY Cha via \citealt{Schwarz-etal-2024}; 
PDS 70 via \citealt{Perotti-etal-2023}; 
TW Hya via \citealt{Henning-etal-2024}; 
T Cha via \citealt{Xie-etal-2025}; 
SZ Cha via \citealt{Espaillat-etal-2023}; 
UX Tau A via \citealt{Espaillat-etal-2024}; 
MY Lup via \citealt{Salyk-etal-2025}; 
RY Lup via \citealt{c2d_Evans+2003}; 
HP Tau via \citealt{Romero-Mirza-etal-2024b};
AS 205 S via \citealt{Olofsson-etal-2009};
DoAr 25 via \citealt{Olofsson-etal-2009}; 
IRAS 04385+2550 via \citealt{Furlan-etal-2006}; 
SR 4 via \citealt{McClure-etal-2010}). 
Qualitatively, DoAr 33, J1615, TW Hya, SY Cha, PDS 70, MY Lup, HP Tau, DoAr 25, and SR 4, each with CO$_2$ and/or C$_2$H$_2$ detections, display elongated, flatter silicate features indicative of grains larger than those in the unprocessed ISM. Of these objects, TW~Hya is notable for hosting an inner disk within $\lesssim$5~au detected by ALMA \citep{Andrews-etal-2016}. While it does show some CO$_2$ emission, it is not as bright as in J1615. In contrast, SZ~Cha, UX~Tau~A, T~Cha, RY~Lup, and GM~Aur, none of which show carbon emission, exhibit more ``amorphous'' silicate features resembling the ISM, produced by smaller, less processed dust grains. These groupings do not hold, however, for IRAS 04385+2550, which displays C$_2$H$_2$ and CO$_2$ with a smooth 10-$\mu$m feature. Additional observations of transitional disks are necessary to evaluate this potential trend between dust processing and carbon visibility.

\section{Summary} \label{conclusions} \label{summary}

We present new JWST spectra and contemporaneous {\halpha} measurements of GM~Aur and J1615, highlighting the carbon-rich chemistry revealed by JWST in the transitional disk of J1615. We report the following results:
\begin{enumerate}
    \item We identify H$_2$O, HCN, C$_2$H$_2$, $^{12}$CO$_2$, $^{13}$CO$_2$, OH, and $^{13}$C$^{12}$CH$_2$ in the JWST-MIRI MRS spectrum of J1615 between $13.6 – 17.7$ $\mu$m. We find that $^{12}$CO$_2$'s \textit{Q}-branch is exceptionally bright, while H$_2$O is comparatively dim. Compared to J1615, the JWST-MIRI MRS spectrum of GM Aur lacks significant molecular emission, with positive detections of only H$_2$O and OH in our modeled wavelength range. 
    \item We use LTE slab models to constrain the column density, temperature, and emitting radius of molecules detected between $13.6 – 17.7$ $\mu$m in J1615 and GM Aur. The derived CO$_2$-to-H$_2$O column density ratio in the disk of J1615 is $N_{CO_2}/N_{H_2O} \sim 2.15$, much higher than measured by previous \textit{Spitzer} surveys of other T Tauri disks \citep{Salyk-etal-2011}. 
    \item We detect a warmer water component between $6.8 - 7.5$ $\mu$m in the disk of J1615 and not in GM Aur, suggestive of a water reservoir closer to J1615's central star.
    \item The detection of C$_2$H$_2$ and CO$_2$ in the disk of J1615 is unusual considering the lower detection rates of these species in transitional disks \citep{JDISC-2025}.
    \item We find evidence of OH prompt emission in both GM Aur and J1615 shortward of 12 $\mu$m, with the former having brighter and more frequent detections. This is suggestive of active H$_2$O photodissociation in the disk of GM Aur due to stronger FUV flux.
    \item We present an SED model of J1615's disk, which is best recreated with larger dust grain sizes and increased levels of crystalline silicates compared to GM Aur. This is indicative of more advanced dust processing/evolution in the disk of J1615.
    \item GM Aur and J1615 show positive detections of [\ion{Ar}{2}] (6.98 $\mu$m), [\ion{Ne}{2}] (12.81 $\mu$m), and [\ion{Ne}{3}] (15.5 $\mu$m). [\ion{Ar}{2}], [\ion{Ne}{2}] and [\ion{Ne}{3}] have all been observed previously in both J1615 and GM Aur \citep{Szulagyi-etal-2012, JDISC-2025}. While [\ion{Ne}{3}] flux has previously been reported in GM Aur \citep{JDISC-2025}, this is the first [\ion{Ne}{3}] flux measurement in J1615 since the upper limit published in \cite{Szulagyi-etal-2012}. The [\ion{Ne}{3}]/[\ion{Ne}{2}] and [\ion{Ne}{2}]/[\ion{Ar}{2}] flux ratios are consistent with X-ray photoevaporation of the disks of GM Aur and J1615.
    \item We measure accretion rates for GM~Aur and J1615 of 2.0$\times$10$^{-8}${\msunyr} and 2.5$\times$10$^{-9}${\msunyr}, respectively, by modeling {\halpha} spectra obtained within 12 hours and 3 hours of the JWST observations.  
    \item We report MIR continuum variability in GM Aur, within the ranges previously seen by \textit{Spitzer}.

\end{enumerate}

J1615 is a carbon-rich transitional disk exhibiting strong molecular emission from hydrocarbons and CO$_2$. Despite having similar spectral types, ages, and dust cavity sizes, GM~Aur and J1615 differ markedly in their mid-infrared molecular emission. We investigate several possible explanations linked to their main physical differences, which include accretion rate, dust grain sizes and crystallinity, and a bright, compact inner disk in the mm. These explanations encompass vertical transport, FUV photodissociation, dust grain opacity, and dust cavities/traps. No single mechanism fully accounts for the observed differences, suggesting that multiple processes may operate simultaneously and highlighting the value of transitional disks as laboratories for studying carbon chemistry. Expanding JWST observations of transitional disks will be crucial for assessing the prevalence and origins of carbon-rich disk chemistry in transitional disks, shedding new light on chemical reactions in the inner disk.


\begin{acknowledgments}
We acknowledge the support of JWST grant GO-01676. We thank M. Colmenares, E. Macias, I. Pascucci, and A. Sellek for insightful discussions. This paper utilizes the D’Alessio Irradiated Accretion Disk (DIAD) code. We acknowledge the important work carried out by Paola D’Alessio, who passed away in 2013. Without her, works like these could not have been achieved.
\\

The JWST data presented in this paper were obtained from the Mikulski Archive for Space Telescopes (MAST) at the Space Telescope Science Institute. The specific observations analyzed can be accessed via \dataset[10.17909/0569-e440]{https://doi.org/10.17909/0569-e440} and \dataset[10.17909/ee96-rs72]{https://doi.org/10.17909/ee96-rs72} for the JWST MIRI MRS observations of GM Aur and RX J1615.3-3255. 
\end{acknowledgments}


\facilities{JWST, CTIO/SMARTS}


\appendix

\section{$\chi^2$  maps} \label{chi2s}

\begin{figure*}
\includegraphics[width=\textwidth]{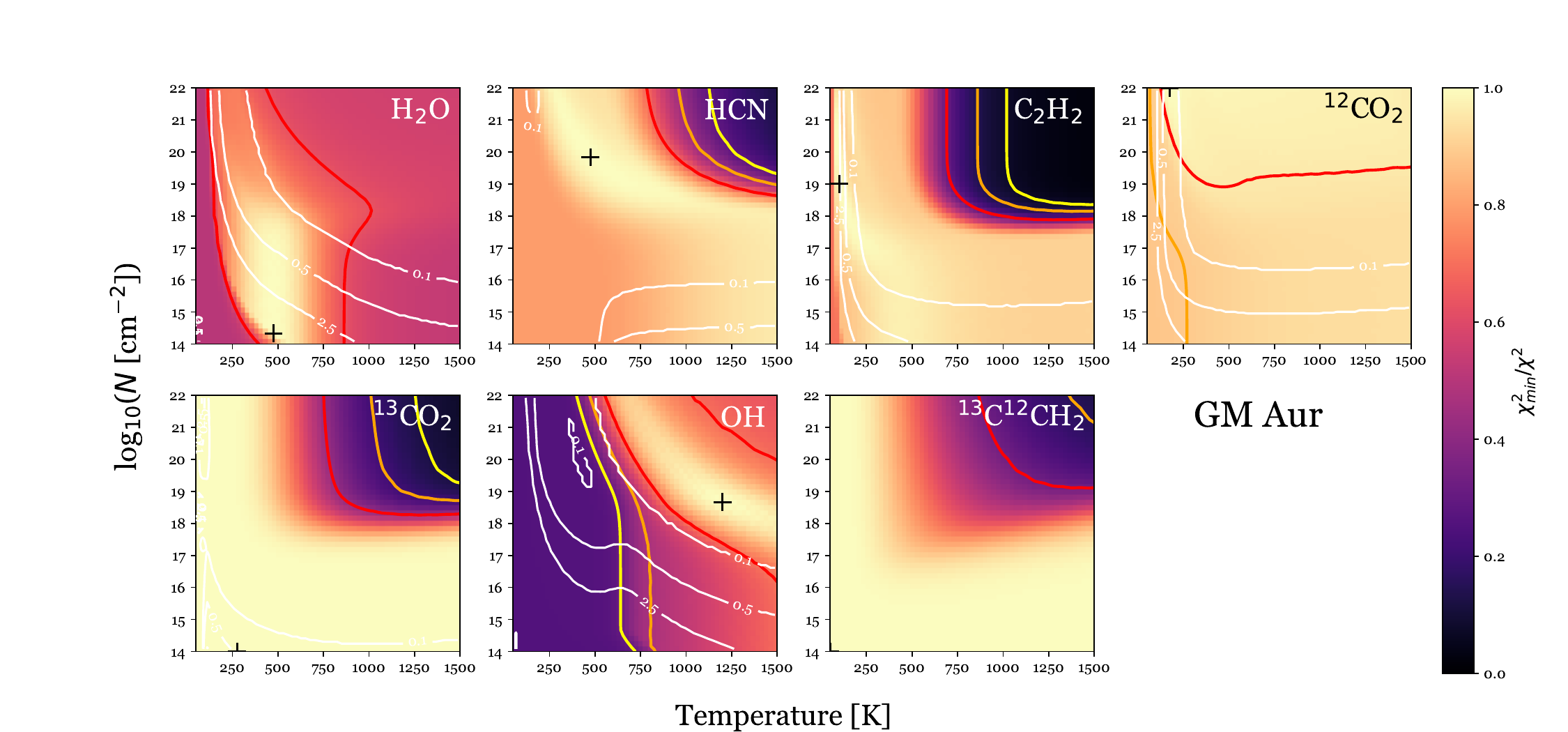}
\includegraphics[width=\textwidth]{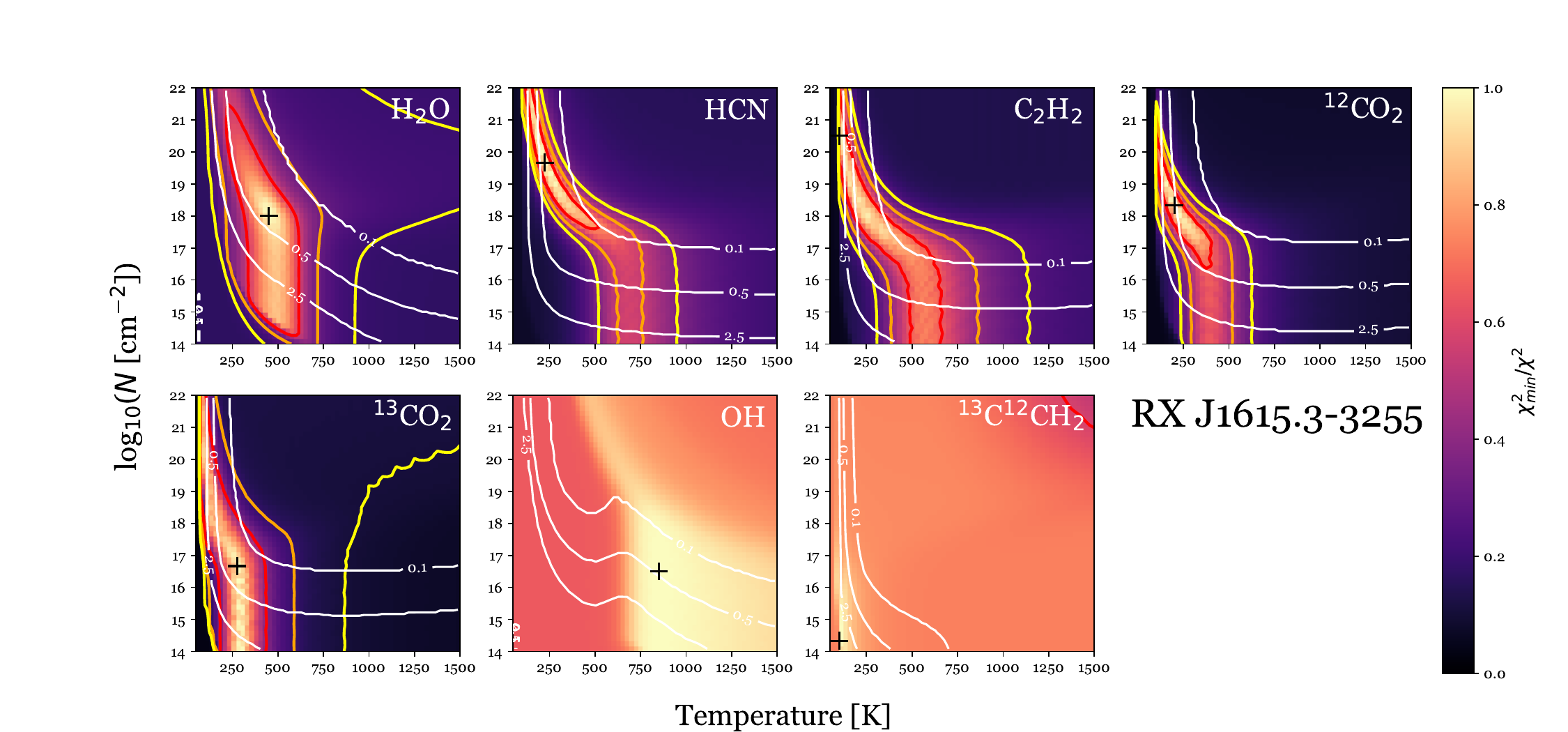}
\caption{The $\chi^2$ maps for H$_2$O, HCN, C$_2$H$_2$, $^{12}$CO$_2$, $^{13}$CO$_2$, OH, and $^{13}$C$^{12}$CH$_2$ from the slab modeling of GM Aur and J1615's spectra between $13.6 - 17.7$ $\mu$m. The color map corresponds to $\chi_{min}^2/\chi^2$. The black crosses are the location of the final best-fit values for column density \textit{N}, temperature \textit{T}, and emitting radius \textit{R} and are located where $\chi_{min}^2/\chi^2 = 1$. The x-axis of each subfigure are all tested temperature values, and the y-axis are all possible column density configurations. The white contour lines are the tested emitting radii, whose values are shown in white text on each map. The red, orange, and yellow contours show the $1\sigma$, $2\sigma$, and $3\sigma$ levels.}
\label{fig:chi2s}
\end{figure*}

\begin{figure*}
    \centering
    \includegraphics[width=\linewidth]{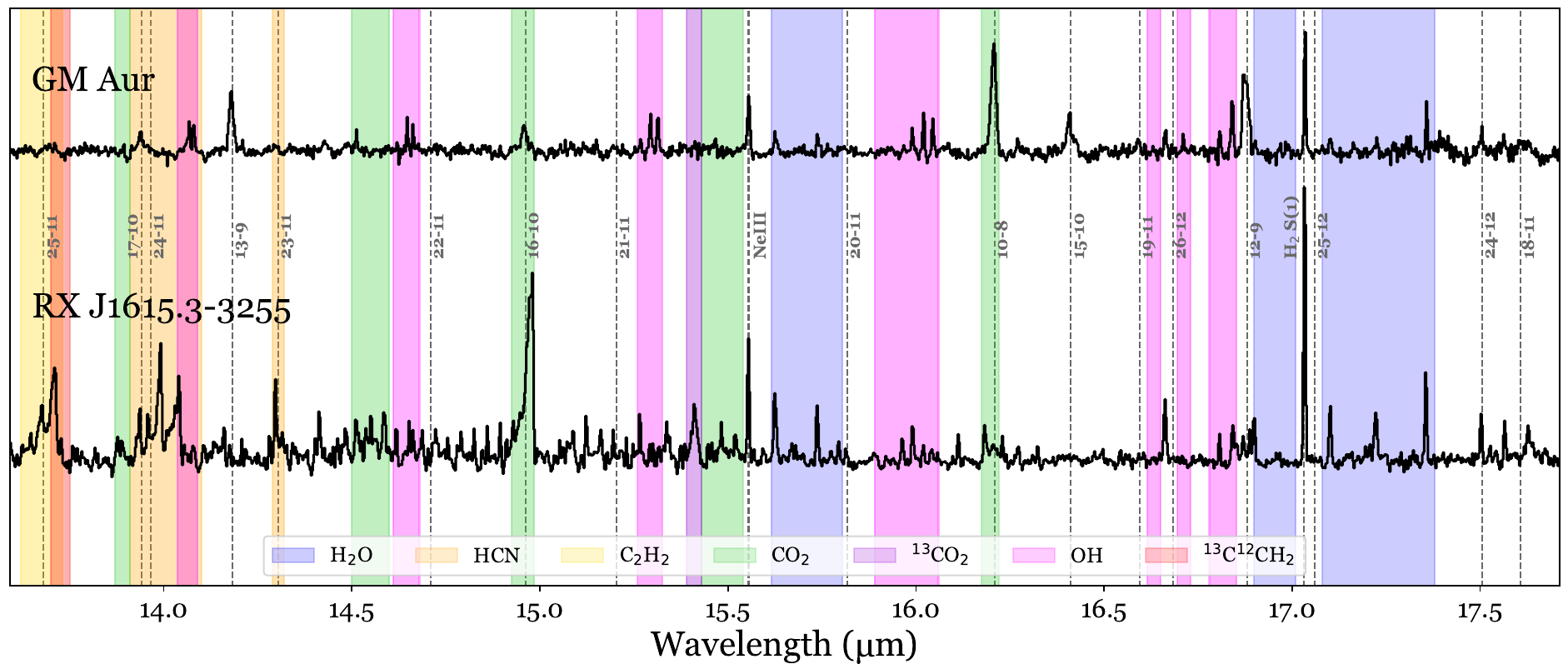}
    \caption{The continuum-subtracted spectra of J1615 and GM Aur (\textit{black}, solid lines) overplotted with the spectral windows for each species (colored regions) used in the slab modeling outlined in Section \ref{slabmodels}. The fitting ranges are chosen to reduce contamination between species and with any potential nearby \ion{H}{1}, H$_2$, and atomic lines (\textit{grey}, dashed lines), while also including emission features useful for constraining the best-fit parameters. Not all lines labeled here are positive detections; see Figure \ref{fig:fullspec} in Section \ref{MIRdata} for a comprehensive view of the observed emission lines.}
    \label{fig:fittingwindows}
\end{figure*}

\begin{figure*}
\includegraphics[width=\textwidth]{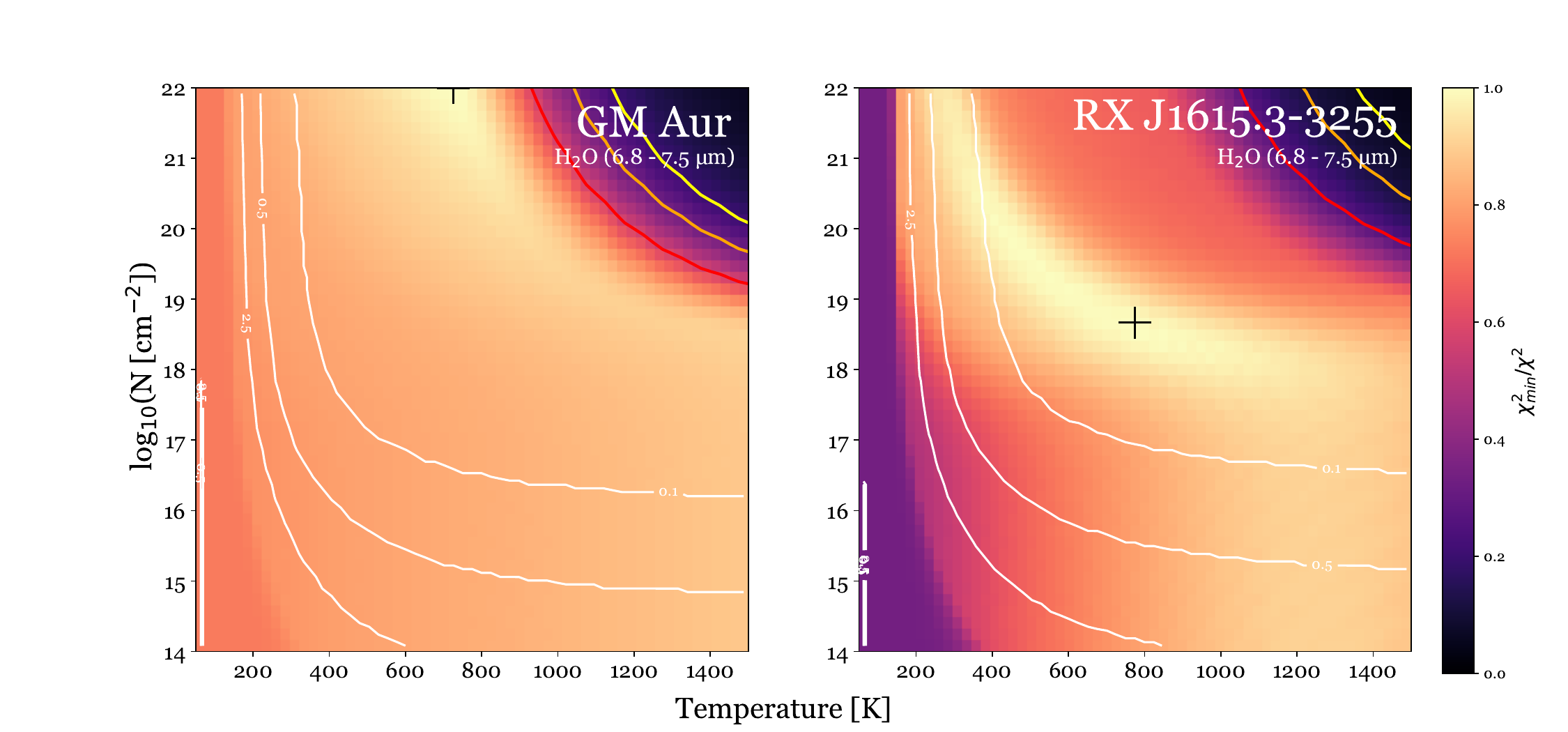}
\caption{The $\chi^2$ maps for H$_2$O as slab modeled in GM Aur and J1615 between $6.8 - 7.5$ $\mu$m. The color map corresponds to $\chi_{min}^2/\chi^2$. The black crosses are the location of the final best-fit values for column density \textit{N}, temperature \textit{T}, and emitting radius \textit{R} and are located where $\chi_{min}^2/\chi^2 = 1$. The x-axis of each subfigure are all tested temperature values, and the y-axis are all possible column density configurations. The white contour lines are the tested emitting radii, whose values are shown in white text on each map. The red, orange, and yellow contours show the $1\sigma$, $2\sigma$, and $3\sigma$ levels.}
\label{fig:chi2s_h2o}
\end{figure*}

In Figures \ref{fig:chi2s} and \ref{fig:chi2s_h2o}, we present the reduced $\chi^2$ maps described in Sections \ref{slabmodels} and \ref{h2o}. The reduced $\chi^2$ is calculated where $N$ is the number of resolution elements in the molecule's fitting window, as shown in Figure \ref{fig:fittingwindows} and the blue horizontal lines in Figure \ref{fig:6-7um_slabs}. We choose these windows in an effort to maximize the molecular features and minimize contamination from other species. $\sigma$ is the standard deviation in the difference between the flux and the continuum level between 15.90 - 15.94 $\mu$m, where we expect little molecular emission. The red, orange, and yellow contours in each $\chi^2$ map represent the 1-, 2-, and 3-$\sigma$ levels determined as $\chi_{min}^2 + 2.3$, $\chi_{min}^2 + 6.2$, and $\chi_{min}^2 + 11.8$ (see \cite{Press-1992} and Equation (6) of \cite{Avni-1976}). 

\section{Photometry} \label{photo}

In Table \ref{table:PhotometrySources}, we list the photometric sources used to guide the SED models derived in Section \ref{sed}.

\begin{table}[]
\hspace{-3em}
\begin{tabular}{lll}
\hline
Label     & Reference       \\
\hline
\hline
GSC2.3    & \cite{GSC2.3_Lasker+2008} \\
GAIA DR3  & \cite{GaiaCollaboration-etal-2023} \\
2MASS     & \cite{2MASS_Cutri+2021} \\
AKARI     & \cite{AKARI-Yamamura-etal-2010} \\
WISE      & \cite{WISE_Cutri+2021} \\
c2d       & \cite{c2d_Evans+2003} \\
AAVSO     & \cite{AAVSO_Henden+2015} \\
IRAS      & \cite{IRAS_Abrahamyan+2015} \\
PANSTARRS & \cite{PANSTARRS_Chambers+2016} \\
Herschel  & \cite{Herschel_Ribas+2017} \\
SCUBA     & \cite{Mohanty-etal-2013} (GM Aur) \\
          & \cite{vanderMarel-etal-2016} (J1615) \\
SMA       & \cite{Andrews-etal-2013} (GM Aur) \\
          & \cite{Andrews-etal-2011} (J1615) \\
ALMA      & \cite{Francis-vanderMarel-2020} (GM Aur) \\
          & \cite{vanderMarel-etal-2016} (J1615) \\
\hline
\end{tabular}
\caption{Sources of the photometry presented in Figure \ref{fig:SEDs}. Entries apply to both targets unless otherwise noted.}
\label{table:PhotometrySources}
\end{table}


\clearpage
\bibliography{rxj-ads-bibtex}
\bibliographystyle{aasjournal}


\end{document}